\documentclass[aps,prd,twocolumn,reprint,nofootinbib,superscriptaddress]{revtex4-1}
\usepackage{CJK}
\usepackage[utf8]{inputenc}
\usepackage[english]{babel}
\usepackage{amssymb,amsmath,latexsym,mathrsfs}
\usepackage{feynmp-auto} 
\usepackage{graphicx}
\usepackage{makeidx}
\usepackage{subfigure}
\usepackage{epsfig}
\usepackage{multirow}
\usepackage{array}
\usepackage{color}
\usepackage{longtable}
\usepackage{multirow}
\usepackage{bigstrut}
\usepackage{hyperref}

\newcommand{\LCDM}{$\Lambda \mathrm{CDM}$}

\newcommand{\eV}{\mathrm{eV}}

\newcommand{\MeV}{\mathrm{MeV}}

\newcommand{\Mpc}{\mathrm{Mpc}}
\newcommand{\neff}{N_\mathrm{eff}}
\newcommand{\geff}{g_{\rm{eff}}}
\newcommand{\zrec}{z_{\nu\mathrm{rec}}}

\newcommand{\om}{\Omega_m h^2}
\newcommand{\omm}{\Omega_m h^{3.4}}

\begin{document}

\title{Cosmological constraints on neutrino self-interactions with a light mediator}
\author{Francesco Forastieri}
\affiliation{Dipartimento di Fisica e Scienze della Terra, Universit\`a degli Studi di Ferrara, via Giuseppe Saragat 1, I-44122 Ferrara, Italy}
\affiliation{Istituto Nazionale di Fisica Nucleare, Sezione di Ferrara, via Giuseppe Saragat 1, I-44122 Ferrara, Italy}
\author{Massimiliano Lattanzi}
\affiliation{Istituto Nazionale di Fisica Nucleare, Sezione di Ferrara, via Giuseppe Saragat 1, I-44122 Ferrara, Italy}
\author{Paolo Natoli}
\affiliation{Dipartimento di Fisica e Scienze della Terra, Universit\`a degli Studi di Ferrara, via Giuseppe Saragat 1, I-44122 Ferrara, Italy}
\affiliation{Istituto Nazionale di Fisica Nucleare, Sezione di Ferrara, via Giuseppe Saragat 1, I-44122 Ferrara, Italy}

\begin{abstract}
If active neutrinos undergo non-standard (``secret'') interactions (NS$\nu$I) the cosmological evolution of the neutrino fluid might
be altered, leaving an imprint in cosmological observables. We use the latest publicly available CMB data from Planck to constrain NS$\nu$I inducing $\nu-\nu$ scattering, under the assumption that the mediator $\phi$ of the secret interaction is very light.
We find that the effective coupling constant of the interaction, $\geff^4 \equiv \langle \sigma v\rangle T_\nu^2$, is constrained
at $< 2.35\times10^{-27}$ (95\% Credible Interval, C.I.), which strengthens  to $\geff^4 < 1.64\times10^{-27}$ when Planck non-baseline small-scale polarization is considered. Our findings
imply that after decoupling at $T\simeq 1\,\MeV$, cosmic neutrinos are free streaming at redshifts $z>3800$, or  $z>2300$ if small-scale polarization is included.  These bounds are only marginally improved when data from geometrical expansion probes are included in the analysis to complement Planck. We also find that the tensions between CMB and low-redshift measurements of the expansion rate $H_0$ and the amplitude of matter fluctuations $\sigma_8$ are not significantly reduced. Our results are independent on the underlying particle physics model as long as $\phi$ is very light. Considering a model with Majorana neutrinos and a pseudoscalar mediator we find that the coupling constant $g$ of the secret interaction is constrained at $\lesssim 7\times 10^{-7}$. By further assuming that the pseudoscalar interaction comes from a dynamical realization of the see-saw mechanism, as in Majoron models, we can bound the scale of lepton number breaking $v_\sigma$ as $\gtrsim (1.4\times 10^{6})m_\nu$.
\end{abstract}

\maketitle

\section{Introduction}
\label{sec:I}
The existence of a cosmic background of thermal relic neutrinos is one of the predictions of the standard cosmological model.
In the early Universe, this cosmic neutrino background (C$\nu$B) is kept in equilibrium with the cosmological plasma by weak interactions. 
Later, when the temperature of the Universe drops below $\sim 1\,\MeV$, the interaction probability per Hubble time becomes negligibly small and neutrinos enter the so-called free-streaming regime, in which they essentially move along geodesics and are coupled to the other species only through the gravitational potentials. This picture is strongly supported by observations; in particular, the prediction for the neutrino density at different times of the cosmic history (most notably at the time of Big Bang nucleosynthesis and of hydrogen recombination), as parameterized by the effective number of degrees of freedom $\neff$, is well in agreement with the value inferred from the abundances of light elements and from observations of the cosmic microwave background (CMB). However, the value of $\neff$ by itself only gives partial information about the free-streaming nature of neutrinos. In fact, some of the effects related to additional relativistic species (e.g. the shift in matter-radiation equality) do not depend on the collisional properties of the fluid. On the other hand, more subtle effects might be sensitive to that, like the phase shift in the photon-baryon acoustic oscillations caused by the ``pull'' of free-streaming species \cite{Bashinsky:2003tk,Follin:2015hya,Baumann:2015rya}. Nowadays cosmological data, in particular cosmic microwave background (CMB) observations from the Planck satellite, provide a unique channel to study extensions of the standard \LCDM, possibly grounded in some new physics beyond the standard model (SM) of particles. In some of these extensions, the collisionless behaviour of neutrinos at $T<1\,\MeV$ is modified, due to some hidden interaction in the neutrino sector. Such non-standard (``secret'') neutrino interactions (NS$\nu$I) could also be related to the mechanism giving origin to the small neutrino masses, like in Majoron models \cite{Chikashige:1980ui,Gelmini:1980re,Schechter:1981cv,Valle:2015pba}. Probes of NS$\nu$I include Z-boson decays \cite{bardin1970nu, bilenky1993invisible, bilenky1999secret}, coherent neutrino-nucleus scattering~\cite{Akimov:2017ade,Barranco:2007tz,deNiverville:2015mwa,Dutta:2015vwa,Liao:2017uzy,Farzan:2018gtr}, supernova cooling \cite{kolb1987supernova, manohar1987limit, dicus1989implications, kachelriess2000supernova, zhou2011comment, jeong2018probing}, neutrinoless double $\beta$ decay \cite{cremonesi2014challenges, dell2016neutrinoless, bilenky2012neutrinoless}, detection of ultra-high energy cosmic neutrinos in neutrino telescopes \cite{Ioka:2014kca,Ng:2014pca,Blum:2014ewa,cherry2016short,Denton:2018aml}, Big-Bang nucleosynthesis \cite{ahlgren2013comment, huang2018observational}, and, last but not least, CMB, the focus of this work \cite{Beacom:2004yd, bell2006cosmological, Friedland:2007vv, basboll2009cosmological, Cyr-Racine:2013jua,archidiacono2014updated, oldengott2015boltzmann,Forastieri:2015paa, Forastieri:2017oma, Oldengott:2017fhy, Lancaster:2017ksf, Kreisch:2019yzn,Barenboim:2019tux}. 

In this work we consider the possibility that the three active neutrinos of the SM have secret self-interactions, and derive constraints on the interaction strength from the Planck 2015 observations of the CMB temperature and polarization anisotropies, complemented by external data.
We focus on the case in which the new interaction is mediated by a very light, possibly massless, particle.
Constraints on this scenario were recently obtained in Refs. \cite{Archidiacono:2013dua,Forastieri:2015paa}, using Planck 2013 data. In particular, in Ref.~\cite{Forastieri:2015paa}, we obtained constraints on the neutrino self-interaction rate by explicitly introducing a collision term in the Boltzmann equation for neutrinos, while other works relied on an effective description of the collisional properties of the fluid. 

In terms of the effective coupling constant $\geff$ used in this paper (see next section for the definition), we found $\geff<2.7 \times 10^{-7}$ (95\% credible interval). Here we update our previous analysis using the latest publicly available data from Planck, that include observations of the CMB polarization on a wide range of angular scales. We also consider a specific particle physics model in which secret neutrino interactions might arise, showing the connection between the Lagrangian of the model and the cosmological collision rate of neutrinos.

The case of neutrino interactions mediated by a heavy particle (so that one effectively deals with a four-point interaction) has been instead considered in Refs. \cite{Cyr-Racine:2013jua,Lancaster:2017ksf}, as well as in Ref.  \cite{Archidiacono:2013dua}. Recently, 
four-point interactions between active neutrinos have been considered as a possible way to solve the tensions between different measurements of $H_0$ and $
\sigma_8$ \cite{Kreisch:2019yzn}. Non-standard interactions among sterile neutrinos in a cosmological setting have instead been studied  in Refs. \cite{Archidiacono:2014nda,Archidiacono:2016kkh,Archidiacono:2015oma,Chu:2015ipa,Forastieri:2017oma,Song:2018zyl} as a possible way to reconcile short-baseline neutrino oscillation anomalies with cosmological data. In this paper we will only consider interactions among active neutrinos.

This paper is organized as follows. In Sec.~\ref{sec:2} we briefly explain how secret interactions are implemented in a cosmological framework. Sec.~\ref{sec:D} describes the datasets used in our analysis, while in Sec.~\ref{sec:R} we present the results of our analysis. We discuss the implication of
our findings for a specific particle physics model in Sec.~\ref{sec:PP}, and we finally draw our conclusions in Sec.~\ref{sec:C}.

\section{Neutrino secret interactions in cosmology}
\label{sec:2}
We consider NS$\nu$I between active neutrinos mediated by a very light particle $\phi$.
Such interactions might arise in extensions of the SM of particle physics, and are possibly related to the origin of neutrino masses \cite{Chikashige:1980ui,Gelmini:1980re,Schechter:1981cv,Valle:2015pba}.
However, in order to cover a wide class of models, we will for the moment avoid specifying the details of the underlying particle physics theory 
and just focus on some general features of the new interactions that impact the cosmological phenomenology. We will anyway 
interpret our results in the framework of a specific particle physics model in Sec.~\ref{sec:PP}.

Denoting with $g$ the dimensionless coupling between neutrinos and the $\phi$ boson, the cross section $\sigma$ for $\phi$-mediated neutrino-neutrino
scattering, $\nu+\nu \to \nu + \nu$, will be of the form $\sigma \approx g^4/E^2$, where $E$ is some relevant energy scale (for example, the energy in the center-of-mass frame).
The thermally-averaged cross section times velocity $\langle \sigma v \rangle$ will then be, for relativistic neutrinos at temperature $T_\nu$:
\begin{equation}
\langle \sigma v \rangle = \xi \frac{ g^4}{T_\nu^2} \, ,
\label{eq:sv}
\end{equation}
where $\xi$ is a numerical factor, whose precise value depends on the underlying particle physics model.
The scattering rate $\Gamma\equiv  n_\nu \langle \sigma v \rangle$ is thus
\begin{equation}
\Gamma = 0.183\times\xi\, g^4 T_\nu \, ,
\end{equation}
where we have taken into account that $n_\nu = (3/2)\times (\zeta(3)/\pi^2)\times T_\nu^3\simeq 0.183\times  T_\nu^3$ for a single neutrino family.
This motivates the following phenomenological description of the neutrino scattering rate:
\begin{equation}
\Gamma = 0.183\times\geff^4 T_\nu \, .
\label{eq:gammabin}
\end{equation}
With this definition \footnote{Note that in our previous work \cite{Forastieri:2015paa} we used a slightly different parameterization, namely $\Gamma = \gamma^4 T_\nu$. In order to compare with results obtained here, values of $\gamma^4$ in \cite{Forastieri:2015paa} should then be divided by a factor $0.183$.}, $\geff\equiv \xi^{1/4} g$ is an effective coupling constant that encloses such details as the precise structure of the underlying theory, the effect of thermal averaging, etc. Given a definite form of the Lagrangian of the theory, this can be remapped, to a good approximation, to a collision rate of the form (\ref{eq:gammabin}), for the purposes of its effect on the evolution of cosmological neutrino perturbations. Seen in another way, the quantity that we are actually constraining is
the (temperature-independent in the high-energy limit) combination $\langle  \sigma v \rangle \, T_\nu^2$.

Given that the expansion rate of the Universe decreases faster than $T$ (since $H\propto T^2$ and $T^{3/2}$ during the radiation- and matter-dominated eras, respectively), it follows that, in presence of a hidden interaction mediated by a light particle, neutrinos, after having decoupled from the primordial fluid at $T\sim 1~\MeV$, become collisional again. The redshift $\zrec$, at which this \textit{recoupling}\footnote{This is actually a misnomer, since neutrinos are still decoupled from the rest of the cosmological plasma. What really happens is that they become collisional again.} happens, depends on the strength of the secret interaction, and can be derived by means of the relation $\Gamma (\zrec) \simeq H(\zrec)$.
\begin{figure}
\begin{center}
\includegraphics[scale=0.28]{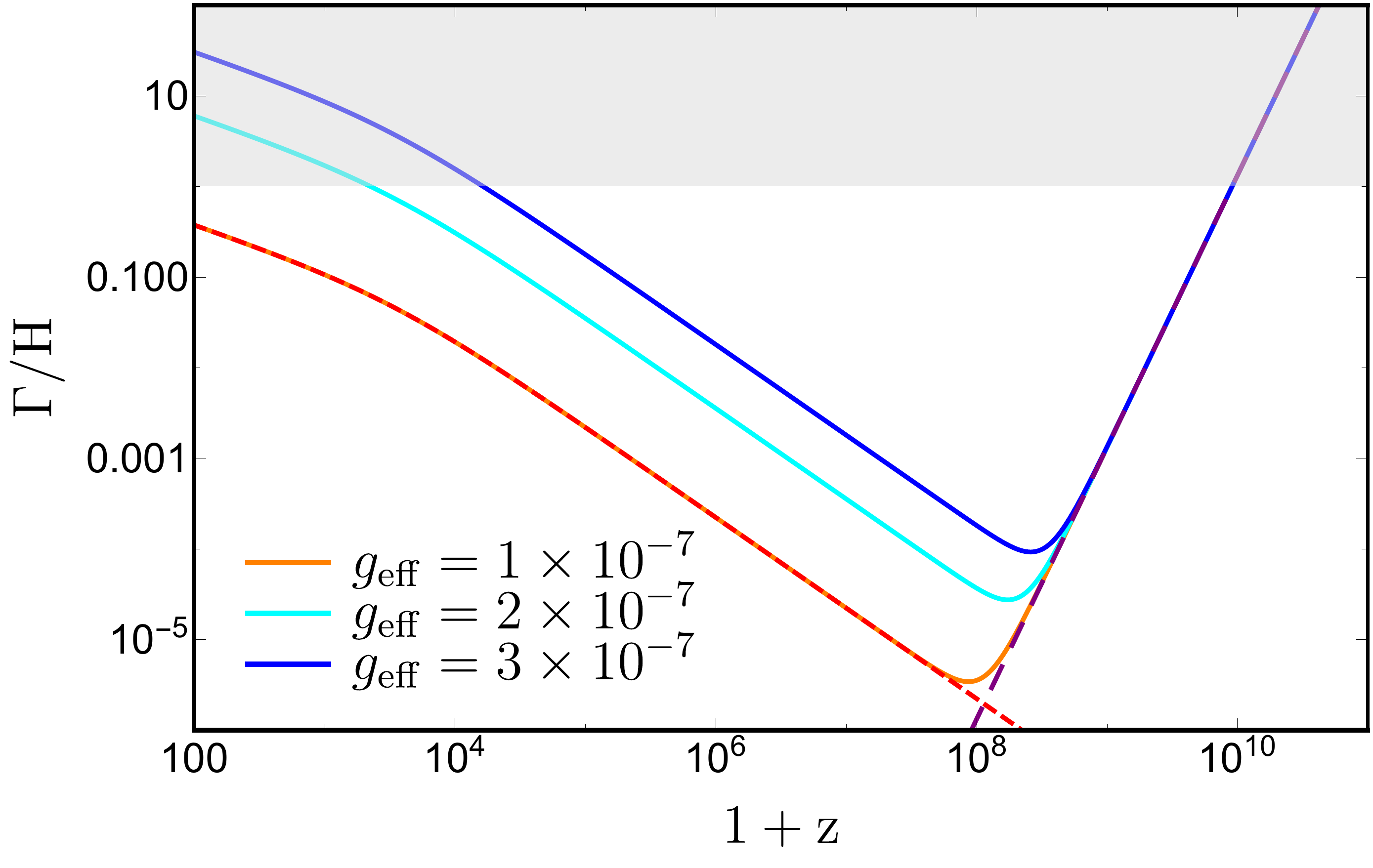}
\caption{Neutrino collision rate $\Gamma$ in units of the expansion rate $H$. The solid lines are drawn considering
the total collision rate, i.e. for both weak processes and $\phi$-mediated interactions, and correspond to $g=\{1,\,2,\,3\}\times 10^{-7}$ from bottom to top.
We also show the usual weak collision rate (purple long dashed line) and the $\phi$-mediated collision rate for $g = 1 \times 10^{-7}$ (red short dashed line).
The gray band shows the region in which the collision rate is larger than the Hubble rate.}
\label{FIG:int}
\end{center}
\end{figure}
To illustrate this, in Fig.~\ref{FIG:int} we show the ratio between the scattering rate $\Gamma$ (also including the contribution from weak processes at high temperatures) and $H$; at early times ($z>10^{10}$) the weak interaction dominates, while later ($z\lesssim10^5 \div 10^3$, for the values considered in the figure), the collisional rate due to the hidden interactions becomes larger than the Hubble expansion rate leading to neutrino recoupling.

Boltzmann codes like \texttt{CAMB} start integrating cosmological perturbations well after weak decoupling, so that neutrinos 
are effectively collisionless at all times of interest in the $\Lambda$CDM model, as well as in many of its more popular extensions. In the scenario considered here, however, once neutrino collisions become relevant again at $z \lesssim \zrec$, their effect on the evolution of perturbations in the neutrino fluid,
and consequently on the cosmological observables, should be taken into account by inserting a suitable collision term in the right-hand side of the Boltzmann equation for the neutrino distribution function $f_\nu$. The exact form of the Boltzmann hierarchy for interacting neutrinos has been derived in Ref.~\cite{Oldengott:2014qra}. Here, as in our previous work \cite{Forastieri:2015paa}, we use the relaxation time approximation and model the collision term
$\hat C[f_\nu]$ as being proportional, through the collision rate $\Gamma$, to the negative of the deviation $\delta f_\nu$ of the distribution function from equilibrium, i.e.
$\hat C[f_\nu]= - \Gamma \delta f_\nu$. 

We choose for simplicity to neglect the effect of non-zero neutrino masses on the cosmological evolution, so that we can approximate neutrinos as massless in \texttt{CAMB} and work with the momentum-integrated version of their Boltzmann hierarchy (see 
Ref. \cite{Ma:1994dv}). Given the precision of the data considered in our analysis (see Sec.~\ref{sec:D}), this is basically equivalent to fix the sum of
neutrino masses $M_\nu$ to $0.06\,\eV$, the minimum value allowed by flavour oscillation experiments. Thus we expect that our limits on $\geff$ should not significantly change if we were to consider massive neutrinos with $M_\nu = 0.06\,\eV$.

\begin{figure}[t]
\centering
\includegraphics[scale=0.315]{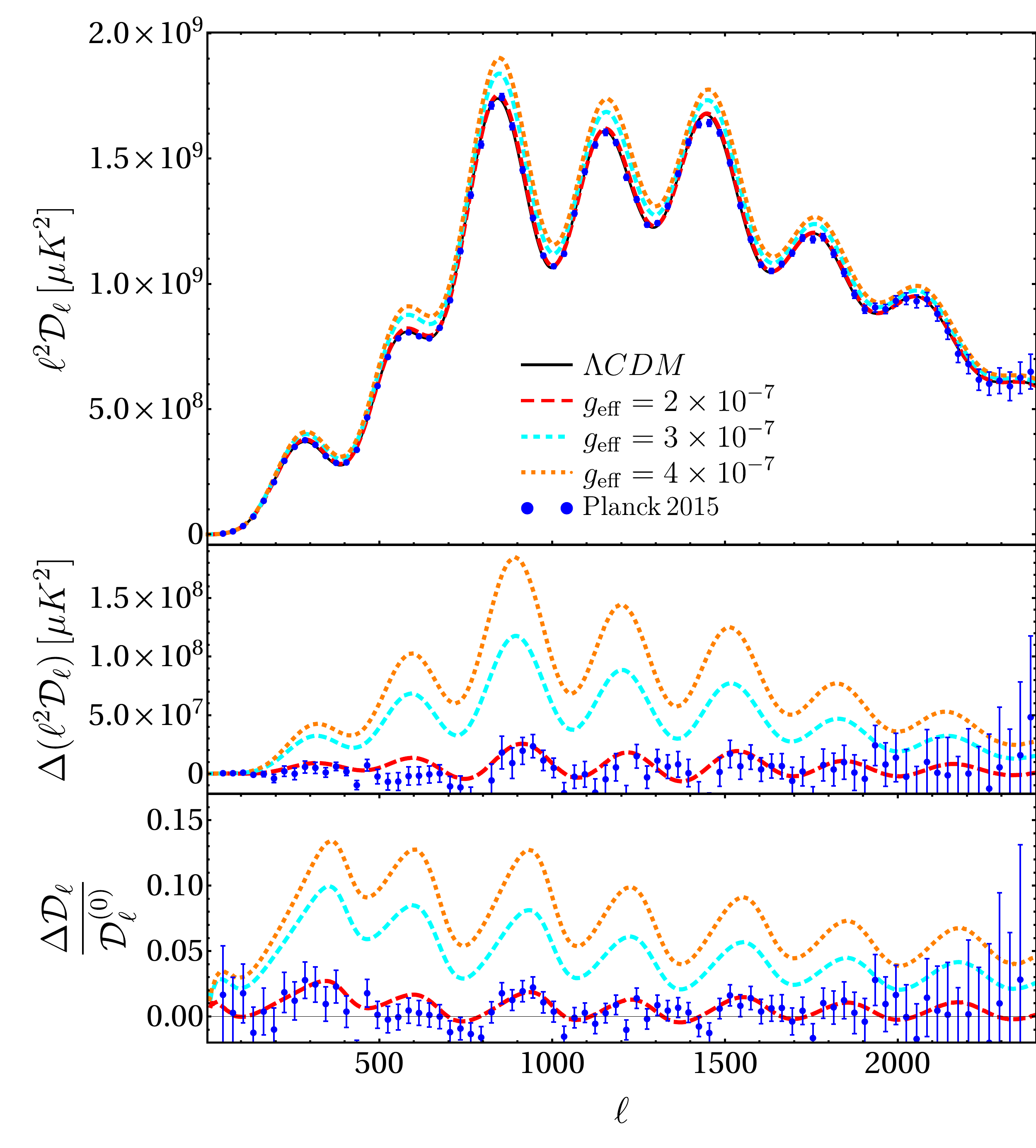}
\caption{Theoretical temperature APS for the \LCDM+$\geff$ model. Note that we are plotting $\ell^2{\mathcal D}_\ell = \ell^3 (\ell+1) C_\ell/2\pi$ to highlight changes at the high multipoles. In the upper panel we show the APS for three different values of the coupling constant, $\geff=\{2,\,3,\,4\} \times 10^{-7}$  (red dashed, cyan dotted, orange dotted curves, respectively). The blue points with error bars are the 2015 Planck data, and the black solid line is the \LCDM\ best-fit ${\mathcal D}^{(0)}_\ell$ to the same data. In the middle panel we plot the differences with respect to \LCDM, $\Delta(\ell^2{\mathcal D}_\ell) = \ell^2({\mathcal D}_\ell-{\mathcal D}^{(0)}_\ell)$. The bottom panel shows the relative difference $\Delta {\mathcal D}_\ell/{\mathcal D}^{(0)}_\ell$.}
\label{FIG:pws}
\end{figure}

\begin{figure}[t]
\includegraphics[scale=0.32]{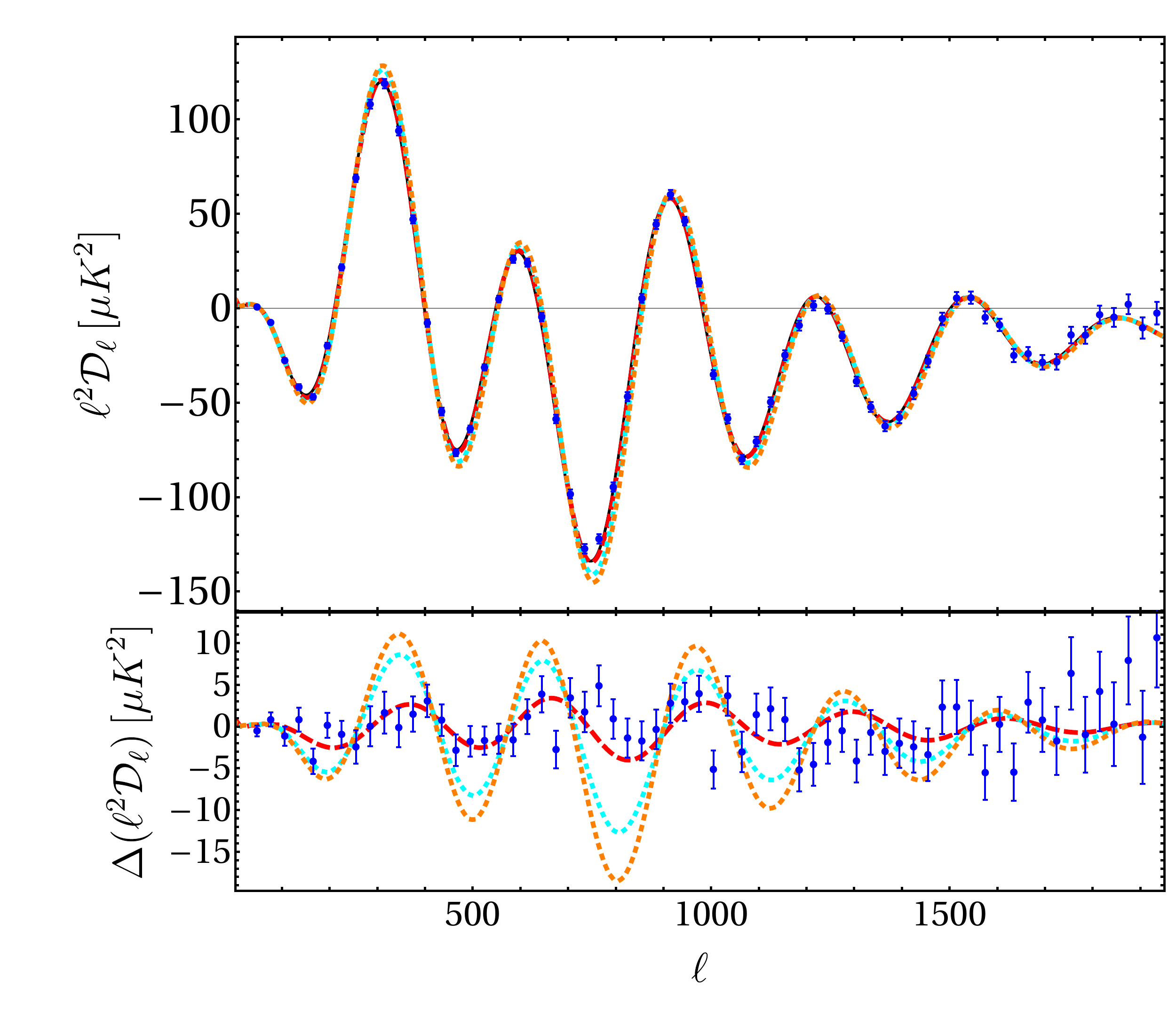}
\caption{Same as Fig.~\ref{FIG:pws}, but for the cross correlation between temperature and $E$-polarization. Note that in this case we do not show relative differences, since these diverge in some points due to the reference spectrum crossing zero.}
\label{FIG:pws2}
\end{figure}

For massless neutrinos, the presence of $\nu - \nu$ scatterings amounts, in the relaxation time approximation,  to modifying the Boltzmann hierarchy as follows:
\begin{widetext}
\begin{subequations}
\begin{align}
& (\ell = 0) \quad \dot{\delta_\nu} = -\frac{4}{3} \theta_\nu - \frac{2}{3} \dot{h} \, ,\\
&(\ell = 1) \quad \dot{\theta}_\nu = k^2 \left(\frac{1}{4} \delta_\nu - \sigma_\nu \right)\, ,\\ 
&(\ell = 2) \quad \dot{\sigma}_\nu = \frac{4}{15} \theta_\nu- \frac{3}{10} k F_{3} + \frac{2}{15} \dot{h} + \frac{4}{5}\dot{\eta} - a \Gamma \sigma_\nu\,  ,\\
&(\ell \ge 3) \quad \dot{F}_{\nu \ell} = \frac{k}{2\ell+1} \Big[ \ell F_{\nu \ell-1} - (\ell+1) F_{\nu \ell+1} \Big]  -a \Gamma F_{\nu \ell}\, .
\end{align}
\label{eq:boltz_hierarchy_int}
\end{subequations}
\end{widetext}
where we use the same notation as Ma \& Bertschinger \cite{Ma:1994dv}, and the monopole and dipole ($\ell=0,1$) of the collision term 
are set to zero, as it follows from the conservation of particle number and momentum.
The $2\leftrightarrow 2$ collisions lead a suppression of the quadrupole ($\ell=2$), i.e. the anisotropic stress $\sigma_\nu$, and of all the highest moments of the distribution function, and to a corresponding enhancement of the monopole and dipole ($\ell=0,1$), i.e. the density and velocity perturbations $\delta_\nu$ and $\theta_\nu$. These changes propagate to the photon distribution, and thus to the CMB spectrum, through the gravitational potentials. In Figs~\ref{FIG:pws} and \ref{FIG:pws2}, we show the effect of neutrino secret interactions on angular power spectrum (APS) of CMB temperature and polarization fluctuations, compared to the standard cosmological model.

Let us now briefly recall how changing the free-streaming nature of neutrinos affects the photon perturbations, that were first clearly established in Ref. \cite{Bashinsky:2003tk}. First of all, since neutrinos and photons are only coupled through the gravitational potentials, decaying rapidly once a perturbation enters the horizon, the effects of neutrino free streaming (or of its absence) on photon perturbations are only relevant at the time of horizon crossing. The impact of neutrino free streaming is twofold. On the one hand, the fast decay of neutrino inhomogeneities after they enter the horizon, and of the associated gravitational potential, damps the photon perturbations. This suppresses the CMB acoustic peaks. On the other hand, the fact that the neutrino velocity exceeds the speed of sound in the baryon-photon fluid generates a distinctive phase shift in the CMB acoustic oscillations. In particular, this ``neutrino pull'' shifts the power spectrum towards larger scales. Both effects are proportional to the fraction of the total energy density provided by free-streaming neutrinos, so they are only relevant for perturbation modes entering the horizon during the radiation dominated era. Altering the collisionless nature of neutrinos by introducing new interactions undoes the effects described above, and will thus appear, at the affected scales, as the combination of a boost and a shift towards larger $\ell$'s (smaller scales) relative to the $\Lambda$CDM case. Since in the model considered here neutrino become noncollisional at $T\simeq 1\,\MeV$ to become collisional again at a later time, we expect to see an impact at scales larger than the horizon at neutrino recoupling (but smaller than the horizon at the time of equality). In Fig.~\ref{FIG:pws}, we show $\ell^2 \mathcal{D}_\ell\equiv \ell^3 (\ell+1) C_\ell/2\pi$ for temperature and for the cross-correlation between temperature and $E$ polarization, for varying values of $\geff$, compared with the best-fit to the Planck 2015 data. The behaviour expected on the ground of the above considerations is indeed observed, especially looking at the lower part of the plots, showing the difference between the NS$\nu$I models and $\Lambda$CDM.

In the discussion so far, we have implicitly assumed that there exists a suitable base of neutrino states in which the interaction is diagonal, and 
that the $\phi$ couples in the same way to all the interaction eigenstates. In other words, if $g_{ij}$ is the matrix of couplings, 
we take $g_{ij} = g \delta_{ij}$. This allows to write a single hierarchy for neutrinos like the one in Eqs.~\ref{eq:boltz_hierarchy_int},
as opposed to three distinct hierarchies with different $\Gamma$'s on the right-hand side. We expect however
that, even if the elements of the coupling matrix are not identical, the limits that we derive can also be regarded as order-of-magnitude constraints for the largest of them.

We also ignore the possibility that the new interaction induces neutrino decay. This amounts to requiring that the off-diagonal elements of $g_{ij}$ vanish in the basis of mass eigenstates. Finally, once neutrinos recouple, a population of $\phi$'s is quickly created by neutrino annihilations, $\nu+\nu \to \phi+\phi$.
Shortly after, the $\phi$ creation is balanced by the inverse reaction $\phi+\phi \to \nu+\nu$ and chemical equilibrium is established. Similarly,
$\nu-\phi$ scatterings drive the system to kinetic equibrium as well. As noted in Refs. \cite{Hannestad:2004qu,Archidiacono:2013dua}, in the limit in which both neutrinos and the new bosons are massless,
this makes Eqs.~(\ref{eq:boltz_hierarchy_int}) also describe the coupled $\nu-\phi$ fluid. The situation is different for massive neutrinos,
since once they become nonrelativistic, the inverse reaction $\phi+\phi \to \nu+\nu$ is suppressed and the neutrino population is rapidly
depleted, leading to a so-called neutrinoless Universe \cite{Beacom:2004yd}.

\section{Method}
\label{sec:D}
We compare the predictions of the model to the CMB observations of the Planck satellite and to additional measurements that constrain
the expansion history of the Universe, like baryon acoustic oscillations (BAO).
In particular, we use the CMB temperature and polarization data publicly released by the Planck collaboration in 2015,
also including the information coming for the lensing reconstruction \cite{Adam:2015rua,Aghanim:2015xee,Ade:2015zua}.
The baseline dataset consists of the TT APS across the whole range of scales measured by Planck ($2\le\ell\le 2500$), denoted $\mathsf{Planck_{15}TT}$
following the conventions of the Planck collaboration papers, together with the low-$\ell$ ($2 \le \ell \le 29$) polarization (\textsf{lowP}). 
For the sake of conciseness, in the following we shall omit to mention the presence of the low-$\ell$ polarization data, but their presence
should be always understood, in any dataset combination. Thus we will refer to the baseline dataset as simply $\mathsf{Planck_{15}TT}$ instead of
$\mathsf{Planck_{15}TT+lowP}$.
The enlarged dataset that also includes the high-$\ell$ ($\ell\ge 30$)
polarization data is similarly denoted as $\mathsf{Planck_{15}TTTEEE}$.

We also consider geometrical information coming various sources: i) measurements of the BAO scale, in particular the BAO results from the 6dF Galaxy Survey \cite{Beutler:2011hx}, from the BOSS DR11 LOWZ
and CMASS samples  \cite{Alam:2016hwk}, and from the Main Galaxy Sample of the Sloan Digital Sky Survey \cite{anderson2014clustering}; 
ii) the Joint Lightcurve Analysis (JLA) supernova sample  \cite{Betoule:2012an}, which is constructed from the
SNLS and SDSS SNe data, joined with several samples of low redshift SNe; iii) the Hubble space telescope data, as
reanalysed in ref.~\cite{Efstathiou:2013via}. The combination of the JLA, BAO and HST dataset will be denoted as ``ext''.

We compute theoretical CMB power spectra using a version of the \texttt{CAMB} code \cite{Lewis:1999bs}, modified
as explained in Sec.~\ref{sec:2}. We derive constraints on the parameters of the model using the Monte Carlo Markov Chain
code \texttt{CosmoMC} interfaced with our modified version of \texttt{CAMB}. 
The constraints are expressed in terms of Bayesian credible intervals (C.I.). The likelihood function associated to the Planck data 
is computed using the code publicly released by the Planck collaboration\footnote{We acknowledge 
the use of the products available at the Planck Legacy Archive (\url{http://www.cosmos.esa.int/web/planck/pla}).}.
We study first a one-parameter extension of the $\Lambda$CDM model in which we add the effective coupling $\geff$
to the six parameters of $\Lambda$CDM. As explained in Sec.~\ref{sec:2}, we always consider massless neutrinos. We take a flat prior on $\geff^4$, and not on $\geff$, since the former is the parameter that enters directly
the perturbation equations (\ref{eq:boltz_hierarchy_int}). We also consider two-parameter extensions of $\Lambda$CDM, in which we also vary one
among the effective number of relativistic species $N_\mathrm{eff}$, the tensor-to-scalar ratio $r$ and the primordial helium abundance $Y_\mathrm{He}$.

\section{Results and discussion\label{sec:R}}
In this section we present our results. We first report constraints on the model parameters, and in particular on the strength of non-standard neutrino interactions (Sec.~\ref{sec:R1}); then we discuss degeneracies among cosmological parameters (Sec.~\ref{sec:R2}).

\subsection{Parameter constraints \label{sec:R1}}
Let us start by considering the minimal extension of the $\mathrm{\Lambda}$CDM model, in which we add the secret interaction strength to the parameter space
of the $\Lambda$CDM model, through the effective coupling constant $\geff$ ($\Lambda$CDM$+ \geff$) defined in Eq. (\ref{eq:gammabin}).
We first show results obtained with the $\mathsf{Planck_{15}TT}$ baseline dataset, and then proceed to add the other datasets described in the previous section.
For the $\mathsf{Planck_{15}TT}$ dataset, results are shown in Tab \ref{tab:1}. We find that the
coupling constant is constrained to be $\geff^4 < 2.9 \times 10^{-27}$. This limit is noticeably tighter, by nearly a factor of 2, with respect to the one obtained in our previous paper \cite{Forastieri:2015paa} using the 2013 Planck results, that reads $\geff^4 < 4.6 \times 10^{-27}$ in terms of the parameterization used in this paper.

\begin{figure}[h!]
\begin{center}
\includegraphics[scale=0.23]{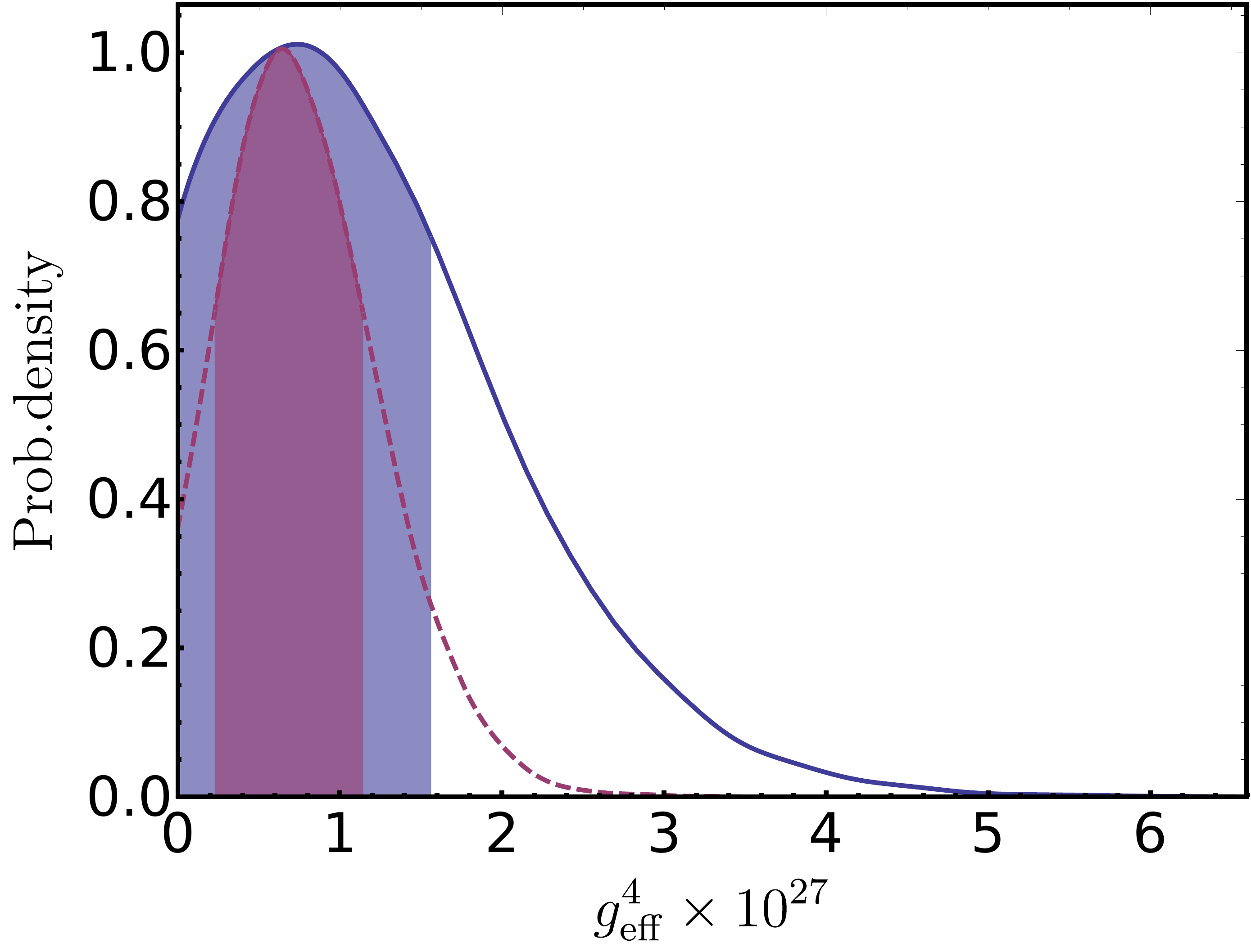}
\includegraphics[scale=0.23]{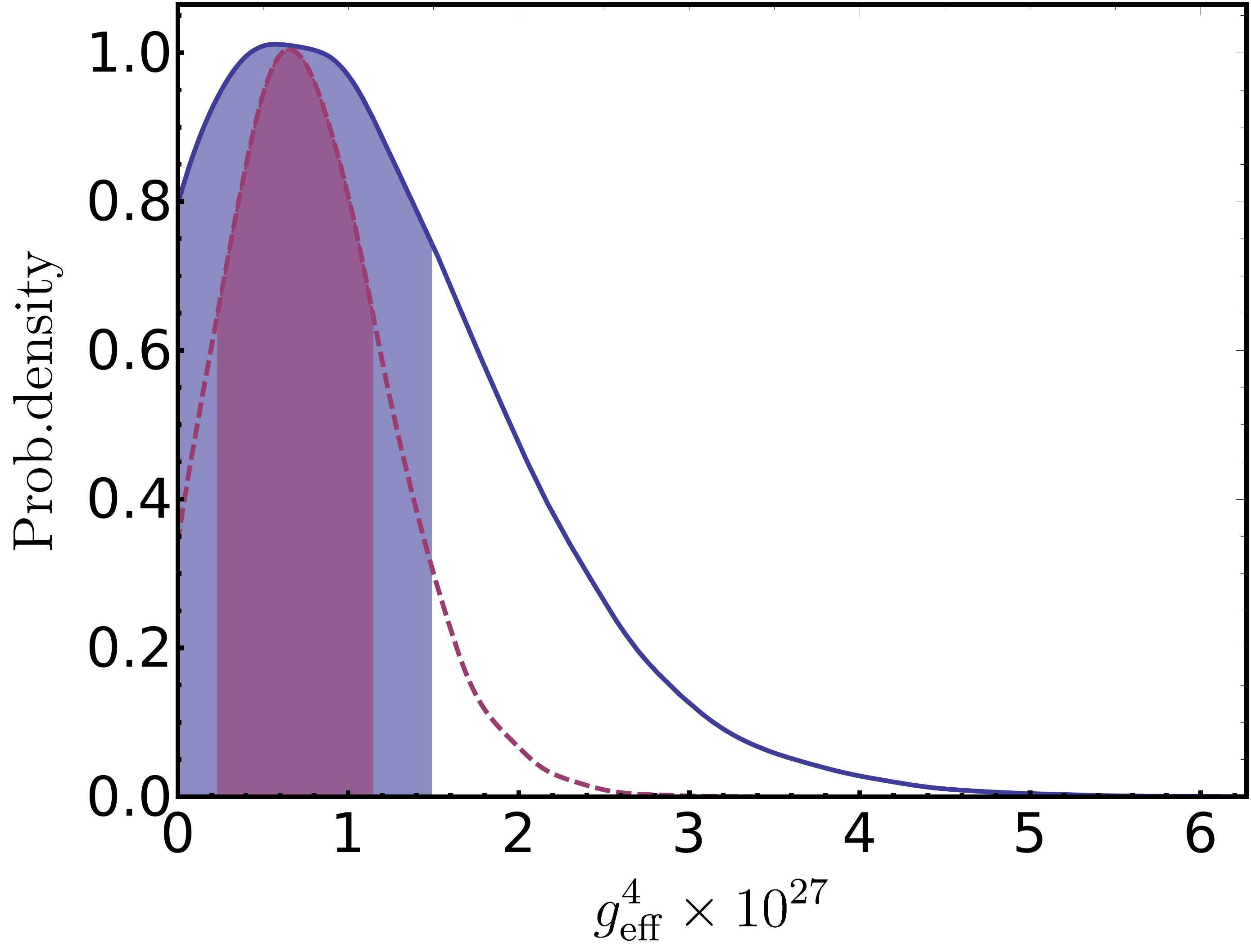}
\includegraphics[scale=0.23]{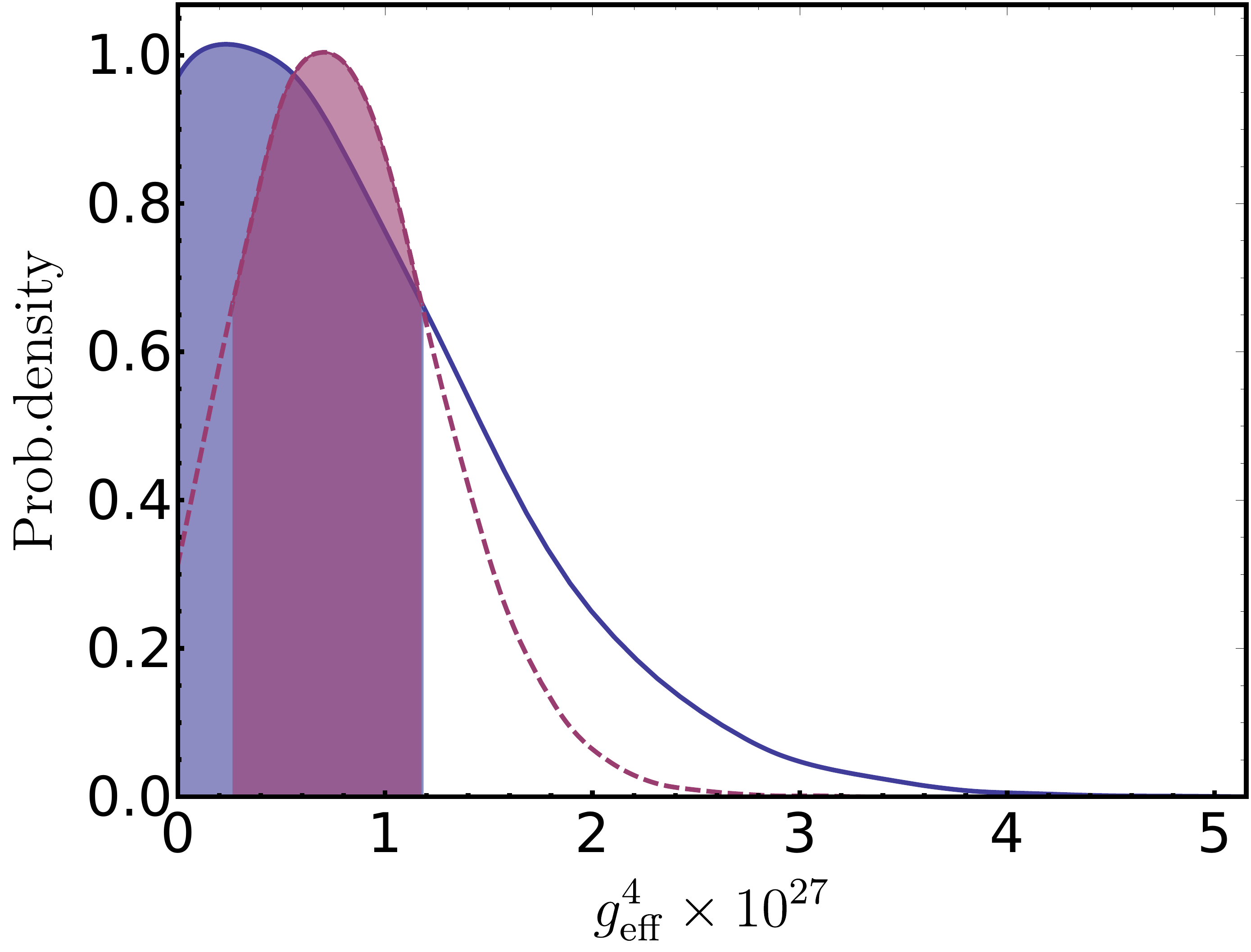}
\end{center}
\caption{One-dimensional posterior probability for the effective parameter $g^4_\mathrm{eff}$ that characterizes the strength of neutrino-neutrino coupling, in the $\Lambda \mathrm{CDM}+g_\mathrm{eff}$ model. The blue (red) curves are obtained using $\mathsf{Planck_{15}TT}$ ($\mathsf{Planck_{15}TTTEEE}$) as the baseline CMB dataset. In the top panel these are the only data considered; in the two lower panels we also add information from external astrophysical datsets [$\mathsf{Planck_{15}TT(TTTEEE)}+$\textsf{ext}, middle panel] or lensing estimates from the CMB 4-point correlation function [$\mathsf{Planck_{15}TT(TTTEEE)}+$\textsf{lensing}, bottom panel]. See text for a more detailed description of the datasets. The shaded areas show 68\% credible intervals.}
\label{FIG:1d}
\end{figure}
\begin{figure}[h!]
\begin{center}
\includegraphics[width=0.4\textwidth]{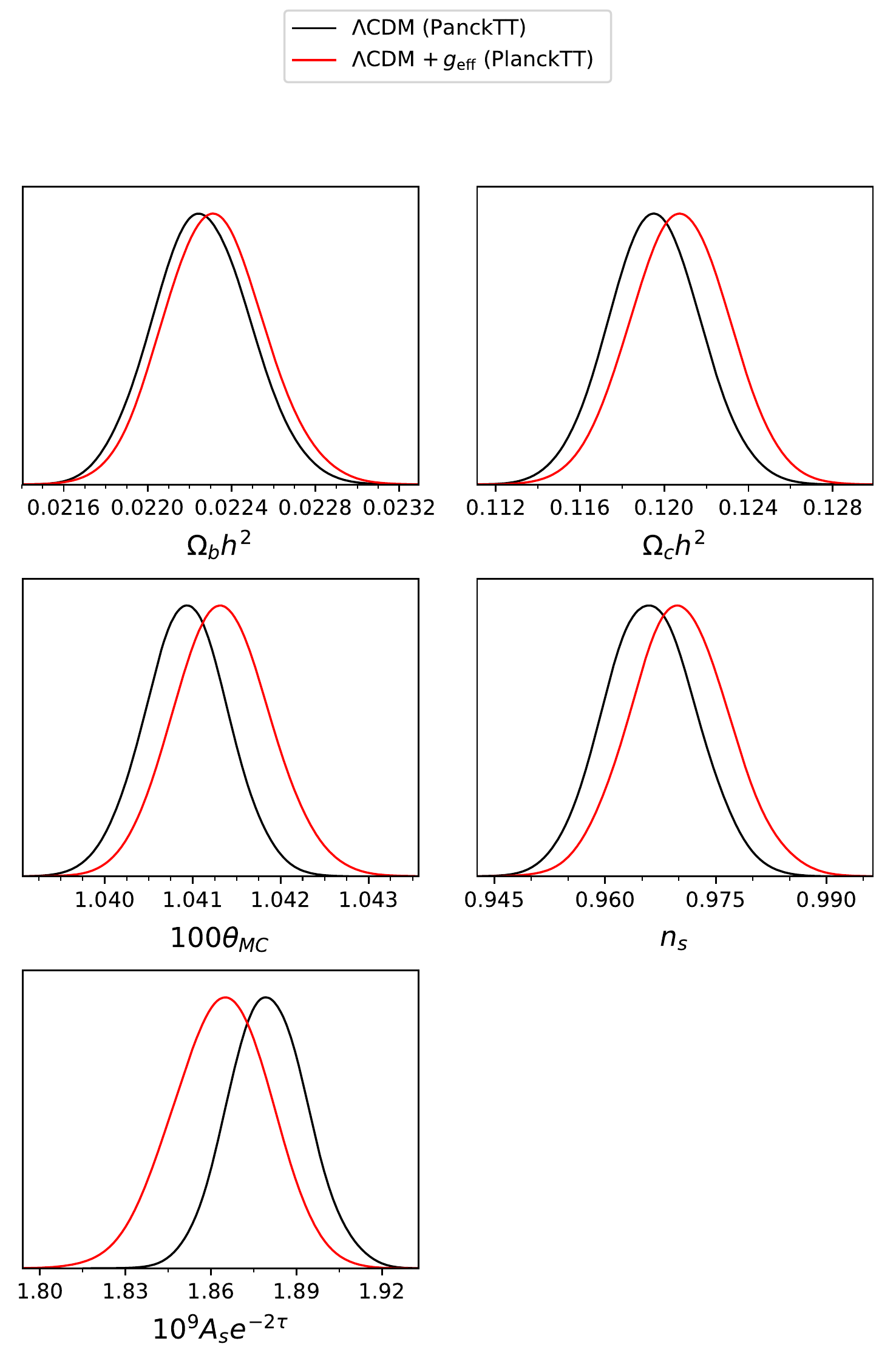}
\end{center}
\caption{One-dimensional marginalized posterior distributions of the base \LCDM\ parameters from $\mathsf{Planck_{15}TT}$,  for the \LCDM\ (black) and \LCDM$+\geff$ (red) models.}
\label{fig:1Dbase}
\end{figure}
When we also consider, in addition to $\mathsf{Planck_{15}TT}$, the astrophysical datasets or the lensing reconstruction, we find (95\% credible intervals) $\geff^4 < 2.8 \times 10^{-27}$ ($\mathsf{Planck_{15}TT}$+ext) and  $\geff^4 < 2.4 \times 10^{-27}$ ($\mathsf{Planck_{15}TT}$+lensing). In both cases results are consistent with respect to those obtained with the baseline dataset. The posterior distributions for $\geff^4$ derived using the $\mathsf{Planck_{15}TT}$,  $\mathsf{Planck_{15}TT}$+ext and $\mathsf{Planck_{15}TT}$+lensing datasets are shown as the blue curves in Fig. \ref{FIG:1d}. In terms of the recoupling redshift $\zrec$, these limits correspond to $\zrec < 5050$ ($\mathsf{Planck_{15}TT}$), $<4750$ ($\mathsf{Planck_{15}TT}$+ext), $<3800$ ($\mathsf{Planck_{15}TT}$+lensing).

The bounds become even tighter when considering the latest public Planck likelihood which includes temperature and polarization data. Results for this case are summarized in Tab.~\ref{tab:2}, and the corresponding posteriors for $\geff^4$ are shown as the red curves in Fig.~\ref{FIG:1d}. We find a mild (roughly at the $1.5\sigma$ level) preference for a non-zero value of the effective coupling constant, that is very stable with respect to the dataset considered. We find $\geff^4 = (0.82^{+0.33}_{-0.60}) \times 10^{-27}$ for $\mathsf{Planck_{15}TTTEEE}$, and nearly identical values when either external astrophysical datasets or lensing is considered. The central value
of the $\geff^4$ posterior corresponds to a redshift of neutrino-neutrino recoupling $z_{\nu rec} \simeq 800$, with a 68\% credible interval $70 < \zrec < 1370$. The 95\% credible interval for $\mathsf{Planck_{15}TTTEEE}$ is $\geff^4 <1.7 \times 10^{-27}$, corresponding to $z_{\nu rec} < 2500$. The improvement observed when the small-scale polarization data is also considered is likely related to the breaking of a degeneracy between $\Omega_m h^2$ and $\geff$ (see next subsection) and on the sharpeness of the polarization peaks that allows to better constrain the phase shift due to non-standard $\nu$ interactions. The results obtained using high-$\ell$ Planck polarization should however not be overinterpreted, since it is possible that a low-level residual systematics is still present in the data \cite{Aghanim:2015xee}.

\begin{figure*}[t]
\begin{center}
\includegraphics[width=\textwidth]{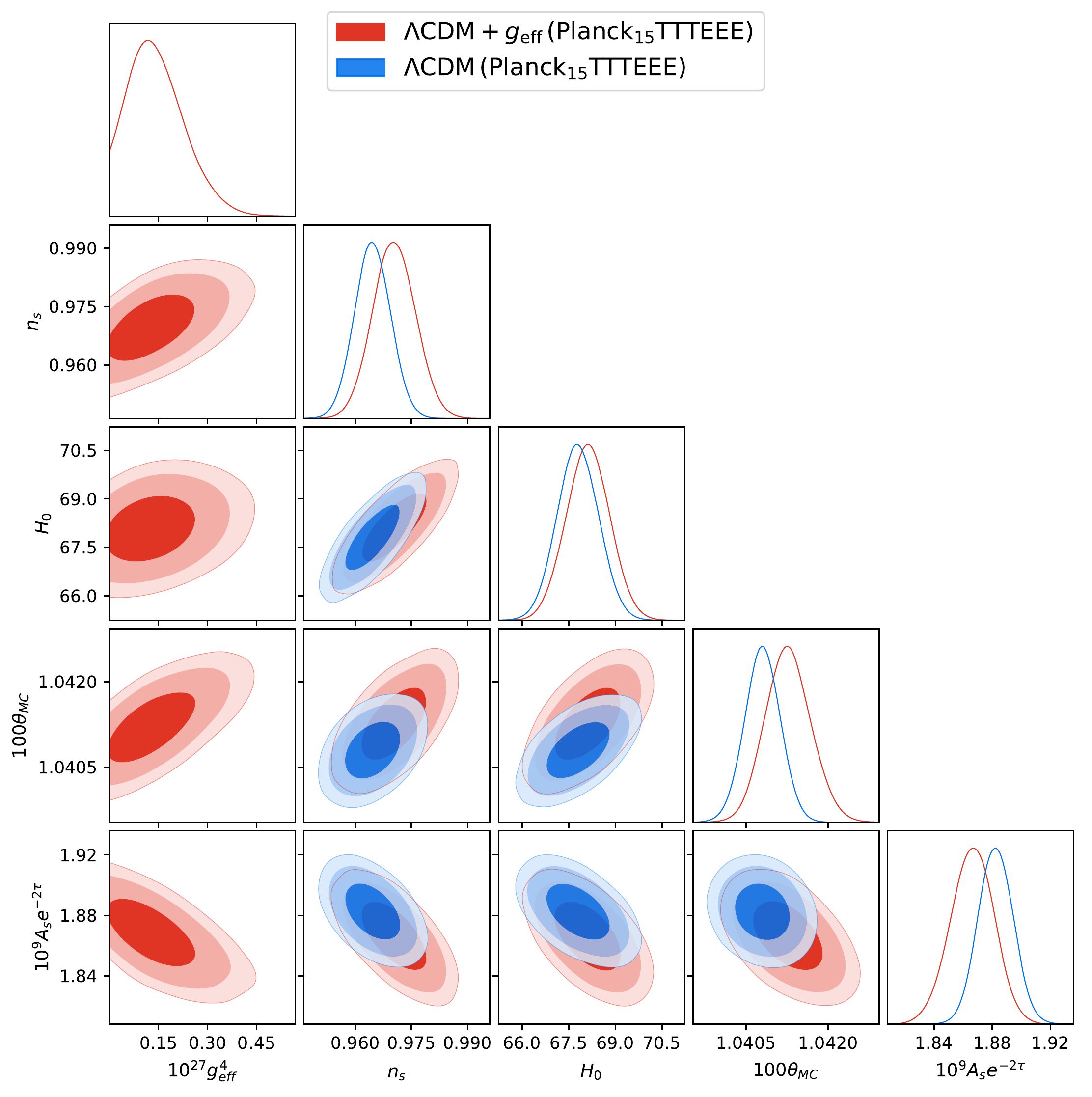}
\caption{One-dimensional posteriors and two-dimensional 68\%, 95\%  and 99\% credible regions for a subset of parameters in the $\Lambda$CDM+$\geff$ (red) and $\Lambda$CDM (blue) models obtained using $\mathsf{Planck_{15}TTTEEE}$ dataset. We consider the following parameters: $\geff^4$, $n_s$, $H_0$, $\theta_s$ and $A_s e^{-2 \tau}$.  }
\label{FIG:tripl}
\end{center}
\end{figure*}

\begin{table*}
\caption{Bayesian credible intervals for the parameters of the $\Lambda \mathrm{CDM}+g_{\mathrm{eff}}$ model obtained using the $\mathsf{Planck_{15}TT}$ (second column), $\mathsf{Planck_{15}TT}+$\textsf{ext} (third column) and $\mathsf{Planck_{15}TT}+$\textsf{lensing} (fourth column) dataset combinations. Unless otherwise noted, we quote 68\% credible intervals.}
\begin{ruledtabular}
\begin{tabular}{cccc}
 & \multicolumn{3}{c}{$\Lambda \mathrm{CDM}+g_\mathrm{eff}$} \\
   \cline{2-4}
\hline
Parameter & $\mathsf{Planck_{15}TT}$ & $\mathsf{Planck_{15}TT}$ & $\mathsf{Planck_{15}TT}$ \\ 
   ~      &                          &      $+$ \textsf{ext}    &  $+$ \textsf{lensing}    \\
\hline
$10^{27}\, g^4_\mathrm{eff}\, \left[95\%\, \mathrm{C.I.} \right]$ & $ < 2.90$ & $ < 2.78$ & $ < 2.35 $ \\
 $\Omega_b h^2$ & $0.02232\pm0.00024$ & $0.02237\pm0.00021$ & $0.02232\pm0.00023$ \\ 
$\Omega_c h^2$ & $0.1207\pm0.0023$ & $0.1200\pm0.0013$ & $0.1190\pm0.0020$ \\ 
$100 \theta_{MC}$ & $1.04134\pm0.00055$ & $1.04142^{+0.00050}_{-0.00055}$ & $1.04143^{+0.00051}_{-0.00056}$ \\ 
$\tau_\mathrm{rei}$ & $0.079\pm0.020$ & $0.082\pm0.018$ & $0.062\pm0.016$ \\ 
$n_s$ & $0.9701\pm0.0066$ & $0.9717\pm0.0054$ & $0.9719\pm0.0063$ \\ 
$\ln(10^{10}A_s)$ & $3.083\pm0.037$ & $3.088\pm0.035$ & $3.047\pm0.030$ \\ 
 \hline 
$10^{7}\, g_\mathrm{eff}\, \left[95\%\, \mathrm{C.I.} \right]$ & $ < 2.33 $ & $ < 2.30  $ & $ < 2.20 $ \\
$z_\mathrm{rec}\, \left[95\%\, \mathrm{C.I.} \right]$ & $< 5050 $& $< 4750  $& $< 3800$\\
$H_0$ [km/s/Mpc] & $67.71^{+0.97}_{-1.10}$ & $68.05\pm0.56$ & $68.35\pm0.93$ \\ 
$\sigma_8$ & $0.845\pm0.015$ & $0.845\pm0.015$ & $0.825\pm 0.009$ \\ 
\end{tabular}
\label{tab:1}
\end{ruledtabular}
\end{table*}

\begin{table*}
\caption{Bayesian credible intervals for the parameters of the $\Lambda \mathrm{CDM}+g_{\mathrm{eff}}$ model obtained using the $\mathsf{Planck_{15}TTTEEE}$ (second column), $\mathsf{Planck_{15}TTTEEE}+$\textsf{ext} (third column) and $\mathsf{Planck_{15}TTTEEE}+$\textsf{lensing} (fourth column) dataset combinations. Unless otherwise noted, we quote 68\% credible intervals.}
\begin{ruledtabular}
\begin{tabular}{cccc}
 & \multicolumn{3}{c}{$\Lambda \mathrm{CDM}+g_\mathrm{eff}$} \\
   \cline{2-4}
\hline
Parameter & $\mathsf{Planck_{15}TTTEEE}$ & $\mathsf{Planck_{15}TTTEEE}$ & $\mathsf{Planck_{15}TTTEEE}$ \\ 
   ~      &                          &      $+$ \textsf{ext}    &  $+$ \textsf{lensing}    \\
\hline
$10^{27}\, g^4_\mathrm{eff}\, \left[95\%\, \mathrm{C.I.} \right]$ & $ < 1.69 $ & $ < 1.64 $  & $ < 1.64 $ \\
$10^{27}\, g^4_\mathrm{eff}\, \left[68\%\, \mathrm{C.I.} \right]$ & $ 0.82^{+0.33}_{-0.60}$ & $ 0.82^{+0.29}_{-0.58}$ & $ 0.82^{+0.29}_{-0.54}$ \\ 
 $\Omega_b h^2$ & $0.02230\pm0.00016$ & $0.02230\pm0.00014$ & $0.02230\pm0.00016$ \\ 
 $\Omega_c h^2$ & $0.1194\pm0.0015$ & $0.1194\pm0.0010$ & $0.1188\pm0.0014$ \\ 
 $100 \theta_{MC}$ & $1.04127\pm0.00041$ & $1.04127^{+0.00037}_{-0.00038}$ & $1.04137^{+0.00039}_{-0.00042}$ \\ 
 $\tau_\mathrm{rei}$ & $0.082\pm0.018$ & $0.083\pm0.017$ & $0.064\pm0.014$ \\ 
 $n_s$ & $0.9704\pm0.0057$ & $0.9705\pm0.0047$ & $0.9714\pm0.0054$ \\ 
 $\ln(10^{10}A_s)$ & $3.091\pm0.034$ & $3.091^{+0.033}_{-0.032}$ & $3.052\pm0.025$ \\
\hline
$10^{7}\, g_\mathrm{eff}\, \left[95\%\, \mathrm{C.I.} \right]$ & $ < 2.03 $ & $ < 2.01 $ & $ < 2.01  $ \\
$z_\mathrm{rec}\, \left[95\%\, \mathrm{C.I.} \right]$ & $< 2500 $& $< 2300 $& $< 2300$\\
$H_0$ [km s$^{-1}$ Mpc$^{-1}$]  & $68.12\pm0.69$ & $68.13\pm0.48$ & $68.38\pm0.67$ \\ 
$\sigma_8$ & $0.844\pm0.013$ & $0.845\pm0.014$ & $0.8262\pm0.0090$ \\ 
\end{tabular}
\end{ruledtabular}
\label{tab:2}
\end{table*}

Then, we enlarge the parameter space adding individually three more parameters: the effective number of extra relativistic degrees of freedom $\Lambda$CDM+$\geff$+$N_{\rm{eff}}$, the primordial helium abundance $\Lambda$CDM+$\geff$+$Y_{\rm{He}}$ and the tensor-to-scalar ratio $\Lambda$CDM+$\geff$+$r$.
Results are reported in Tab.~\ref{tab:3}, and are very similar to what we found in the $\Lambda$CDM+$\geff$ model. Using
the $\mathsf{Planck_{15}TT}$ likelihood, we find $\geff^4 < 3.1 \times 10^{-27}$ ($\Lambda$CDM+$\geff$+$N_{\rm{eff}}$), $3.0 \times 10^{-27}$ ($\Lambda$CDM+$\geff$+$r$), $3.4 \times 10^{-27}$ ($\Lambda$CDM+$\geff$+$Y_{\rm{He}}$), indicating that the constraints on $\geff$ are quite robust in the models considered. 
When using $\mathsf{Planck_{15}TTTEEE}$ data, we still find a $\sim 1\sigma$ preference for non-zero values of the effective coupling constant at $\geff^4 \simeq 0.82 \times 10^{-27}$. The estimates of the additional parameters are also quite stable with respect to the presence of $\geff$: using $\mathsf{Planck_{15}TTTEEE}$ data we obtain  $N_{\rm{eff}}=3.09^{+0.43}_{-0.41}$, $Y_{\rm{He}}=0.247^{+0.027}_{-0.029}$ and $r<0.12$ (95\% credible intervals).
These constraints are very close and only slightly degraded with respect to the ones that can be obtained in the corresponding one parameter extensions of $\Lambda$CDM \cite{Ade:2015xua}.

\begin{table*}
 \caption{Bayesian credible intervals for the parameters of the $\Lambda \mathrm{CDM}+g_\mathrm{eff}+N_\mathrm{eff}$ and $\Lambda \mathrm{CDM}+g_\mathrm{eff}+r$ models from the analysis of the $\mathsf{Planck_{15}TT}$ and $\mathsf{Planck_{15}TTTEEE}$ datasets. We quote $68\%$ credible intervals, except for upper bounds, which are $95\%$.}
\begin{center}
\begin{ruledtabular}
\begin{tabular}{ccccccc}
 & \multicolumn{2}{c}{$\Lambda \mathrm{CDM}+g_\mathrm{eff}+N_\mathrm{eff}$} & \multicolumn{2}{c}{$\Lambda \mathrm{CDM}+g_\mathrm{eff}+r$} & \multicolumn{2}{c}{$\Lambda \mathrm{CDM}+g_\mathrm{eff}+Y_\mathrm{He}$} \\
   \cline{1-7}
Parameter  & $\mathsf{Planck_{15}TT}$   & $\mathsf{Planck_{15}TTTEEE}$    & $\mathsf{Planck_{15}TT}$ &  $\mathsf{Planck_{15}TTTEEE}$ &  $\mathsf{Planck_{15}TT}$ &  $\mathsf{Planck_{15}TTTEEE}$ \\
\hline 
$10^{27}\, g^4_\mathrm{eff}$ $\left[95\%\, \mathrm{C.I.} \right]$ & $ < 3.10 $ & $ < 1.70$ & $ <2.95 $ & $ < 1.70$ & $ < 3.44$ & $ < 1.70$ \\
$10^{27}\, g^4_\mathrm{eff}$ $\left[68\%\, \mathrm{C.I.} \right]$ & / & $ 0.82^{+0.33}_{-0.60}$ & / & $0.82^{+0.30}_{-0.60}$ & / & $ 0.82^{+0.33}_{-0.60}$\\
$\Omega_b h^2$ & $0.02236^{+0.00036}_{-0.00040}$ & $0.02234\pm0.00026$ & $0.02233\pm0.00023$ & $0.02230\pm0.00016$ & $0.02228^{+0.00030}_{-0.00034}$ & $0.02232\pm0.00022$  \\ 
$\Omega_c h^2$ & $0.1212\pm0.0040$ & $0.1200\pm0.0032$ & $0.1206\pm0.0023$ & $0.1193\pm0.0015$ & $0.1211\pm0.0027$ & $0.1194\pm0.0015$ \\ 
$100 \theta_{MC}$ & $1.04132^{+0.00062}_{-0.00069}$ & $1.04125\pm0.00049$ & $1.04140^{+0.00054}_{-0.00059}$ & $1.04132\pm0.00042$& $1.04122^{+0.00082}_{-0.00094}$ & $1.04134\pm0.00064$ \\ 
$\tau_\mathrm{rei}$ & $0.081^{+0.021}_{-0.024}$ & $0.084\pm0.019$ & $0.078\pm0.019$ & $0.081\pm0.018$ & $0.078^{+0.020}_{-0.022}$ & $0.083\pm0.019$  \\ 
$n_s$ & $0.972\pm0.016$ & $0.972\pm0.011$ & $0.972\pm0.007$ & $0.972\pm0.006$& $0.968^{+0.010}_{-0.013}$ & $0.9713\pm0.0084$ \\ 
$\ln(10^{10}A_s)$ & $3.088^{+0.047}_{-0.051}$ & $3.095^{+0.039}_{-0.040}$ & $3.080\pm0.037$ & $3.088\pm0.034$ & $3.079^{+0.041}_{-0.045}$ & $3.093\pm0.037$ \\ 
$N_\mathrm{eff}$ $\left[95\%\, \mathrm{C.I.} \right]$  &  $3.09^{+0.64}_{-0.59}$ & $3.09^{+0.43}_{-0.41}$ & / & / & / & /  \\
$r$ $\left[95\%\, \mathrm{C.I.} \right]$ & / & / &$ < 0.13$ & $ < 0.12$ & / & /  \\
 $Y_{\rm{He}}$  $\left[95\%\, \mathrm{C.I.} \right]$ & / & / & / & / & $0.24^{+0.036}_{-0.04} $ & $0.247^{+0.027}_{-0.029}$ \\ 
\hline
$10^{7}\, g_\mathrm{eff}$ $\left[95\%\, \mathrm{C.I.} \right]$  & $<2.35$  & $<2.05$  & $2.33$ &   $<2.05$ & $ <2.42$ & $ 2.03$  \\
$z_\mathrm{rec}$ $\left[95\%\, \mathrm{C.I.} \right]$ & $<5300$ & $<2400$ & $<5000$ & $<2400$ & $6000$ & $2400$ \\
$H_0$ [km/s/Mpc] & $68.1^{+2.7}_{-3.0}$ & $68.5\pm1.8$ & $67.80\pm0.98$ & $68.20\pm0.69$ & $67.5^{+1.2}_{-1.4}$ & $68.19\pm0.79$\\ 
$\sigma_8$ $\left[95\%\, \mathrm{C.I.} \right]$& $0.848^{+0.035}_{-0.031}$ & $0.847\pm0.03$ & $0.844\pm0.032$ & $0.843\pm0.03$ & $0.844\pm 0.032$ & $0.845\pm0.03$\\
\end{tabular}
 \label{tab:3}
\end{ruledtabular}
 \end{center}
\end{table*}

\subsection{Parameter degeneracies \label{sec:R2}}
In this section we discuss degeneracies among parameters in the model under consideration, together with shifts in parameter estimates with respect to the standard \LCDM\ model. To this purpose, we compare our results with those of control runs with $\geff=0$ \footnote{Note that this is different from the usual base \LCDM\, i.e. the minimal model analyzed in the Planck parameters paper \cite{Ade:2015xua}, since we are neglecting neutrino masses. In any case, if we use as reference values those reported in Ref.~\cite{Ade:2015xua} for base \LCDM\, our conclusions do not change significantly, indicating that neutrino masses are only a subleading effect for what concerns parameter shifts.}. 
Let us start from the $\mathsf{Planck_{15}TT}$ dataset. The one-dimensional posteriors for the base parameters in the two models are shown in Fig.~\ref{fig:1Dbase}. We observe shifts between $\sim~0.5$ and $1.1~\sigma$ in $A_s e^{-2\tau}$, $\theta_s$, $n_s$, and $\Omega_c h^2$ (in order of decreasing magnitude of the shift), and a smaller change ($0.25\sigma$) in $\Omega_b h^2$. In particular, we find
\begin{multline}
\Delta \Big\{\Omega_b h^2,\,\Omega_c h^2,\,\theta_s,\, A_s e^{-2\tau}, \, n_s   \Big\} =\\ = \{0.26,\,0.55,\,0.85,\,-1.14,\,0.64\} \quad \mathsf{(Planck_{15}TT)},
\end{multline}
where the deviation $\Delta$ for a parameter is computed as the difference between the mean value in \LCDM\ and the one in \LCDM+$\geff$, in units of the \LCDM\ uncertainty. Note that we report the shift in $A_s e^{-2\tau}$, instead than in terms of $\ln\left[10^{10} A_s\right]$ and $\tau$ separately, since these are misleadingly small ($\sim0.1\sigma$) due to the larger uncertainty associated to the two individual parameters. In contrast, the combination $A_s e^{-2\tau}$ is much more precisely constrained. We also look at the shift in $\omm$, that is known to be a proxy for the angular size of the sound horizon at recombination \cite{Percival:2002gq}, finding $\Delta(\omm) = 0.51$.
For what concerns parameter correlations, we find that $\geff$ is strongly (correlation coefficient $\gtrsim 0.5$) correlated with $\Omega_m h^{3.4}\,(+0.70)$, $A_s e^{-2\tau}\,(-0.57)$ and $\theta_s\, (+0.48)$, and mildly ($0.25 \lesssim$ correlation coefficient $\lesssim 0.5$) correlated with $\Omega_m h^2\, (+0.38)$, $\Omega_c h^2\, (+0.34)$, $n_s\, (+0.29)$ and $\Omega_b h^2\, (+0.23)$. These correlations nicely reflect the magnitude of the shifts reported above.

Our interpretation of the parameter shifts and correlations in the \LCDM$+\geff$ model is as follows. The large positive shift in $\theta_s$ is related to the absence, in the \LCDM$+\geff$ model, of the phase shift induced by neutrino free-streaming, as described in the previous section: starting from a good \LCDM\ fit to the data, NSI move the spectrum towards smaller scales, and a larger value of $\theta_s$ is required to move it back close to the original model. The shifts in $n_s$, $\Omega_c h^2$ and $A_s e^{-2\tau}$ are instead related to the boost in the angular power spectrum provided by neutrino scatterings. This can be canceled around the first peak by decreasing the primordial amplitude of scalar fluctuations and/or increasing the total matter density. However, since this boost is smaller at high multipoles as seen in Fig.~\ref{FIG:pws}, these will end up being suppressed and can be moved back by increasing the spectral index. We have verified that this sequence of parameter shifts still leaves a small phase shift in the damping tail at $\ell \gtrsim 1000$, that is eliminated by an upward shift in $\Omega_b h^2$.

When we instead use the $\mathsf{Planck_{15}TTTEEE}$ dataset, we find that the shifts in $\theta_s,\,A_s e^{-2\tau},\,n_s$ are increased:
\begin{multline}
\Delta \Big\{\Omega_b h^2,\,\Omega_c h^2,\,\theta_s,\, A_s e^{-2\tau}, \, n_s   \Big\} =\\ = \{0.31,\,-0.27,\,1.5,\,-1.3,\,1.2\} \mathsf{(Planck_{15}TTTEEE)},
\end{multline}
while for $\omm$ we find $\Delta(\omm) = 0.89$. We note that the upward shift in the angle subtended by the sound horizon at recombination reaches the $1.5\sigma$ level. We have also computed parameter correlations, and found strong correlation between $\geff$ and $\Omega_m h^{3.4}\,(+0.66)$, $\theta_s\, (+0.60)$, $A_s e^{-2\tau}\,(-0.57)$, $n_s\, (+0.49)$. On the other hand, we find that the correlations of $\geff$ with $\Omega_c h^2$, $\Omega_b h^2$ and $\Omega_m h^2$ basically disappear, all becoming $\sim -0.1$. We interpret this latter fact in terms of the different effect of a change in the total matter density $\Omega_m h^2$ on the temperature and polarization anisotropies, as discussed in Ref.~\cite{Galli:2014kla}. In fact, while temperature anisotropies are affected both by the decay of gravitational potentials before matter-equality, and by the early-integrated Sachs-Wolfe effect, only the former impacts on polarization anisotropies. This allows to break degeneracies involving $\Omega_m h^2$ (e.g. the $\Omega_m h^2-n_s$ degeneracy in standard \LCDM)  through the inclusion of polarization data. We argue that this is what breaks the $\Omega_m h^2 - \geff$ degeneracy in the model under consideration. Given that changing the matter density is not an option anymore to cancel the effect of $\geff$ on the height of the peaks, the correlation with $A_s e^{-2\tau}$ and $n_s$ is increased. The larger shift in $\theta_s$ is instead mostly due to the smaller uncertainty in its determination, associated to the increased sharpness of the polarization peaks \cite{Galli:2014kla}.

We conclude this section by discussing the shifts in the Hubble constant $H_0$ and in the amplitude of matter fluctuations $\sigma_8$. Note that our reference values are obtained in a \LCDM\ model with massless neutrinos, so they differ from the values quoted in the Planck 2015 parameters paper \cite{Ade:2015xua}; for example, with $\mathsf{Planck_{15}TT}$ we find $H_0=(67.9\pm 1)\,\mathrm{km}\ \mathrm{s}^{-1} \Mpc^{-1}$ and $\sigma_8=0.841\pm 0.015$ for \LCDM.
  Using $\mathsf{Planck_{15}TT}$, we thus find shifts of $-0.20\sigma \, (H_0)$ and $+0.27\sigma\, (\sigma_8$),
while the correlation coefficients with $\geff$ are $-0.14\, (H_0)$ and $0.18\,(\sigma_8)$. Using instead $\mathsf{Planck_{15}TTTEEE}$, we find shifts of $+0.46\sigma \, (H_0)$ and $+0.14\sigma\, (\sigma_8$),
while the correlation coefficients with $\geff$ are $+0.23\, (H_0)$ and $0.08\,(\sigma_8)$. In this case, the $H_0$ shift is in the right direction to alleviate the tension between the cosmological and astrophysical measurements of this parameter \cite{Riess:2016jrr}, but the magnitude of the shift is too small to reduce the tension significantly. The shift in $\sigma_8$, on the other hand, is in the ``wrong'' direction (for both datasets), in the sense that it goes towards exacerbating the tension between high- and low-redshift probes of the clustering amplitude. Even in this case, however, the shift is marginal.

The shifts in $H_0$ can be understood by recalling that CMB observations independently constraint $\Omega_m h^2$ (through the redshift of matter-radiation equality) and $\Omega_m h^{3.4}$ (through the angular size of the sound horizon at recombination \cite{Percival:2002gq}). It is straightforward to write the relative change in $h$ induced by varying $\Omega_m h^2$ and $\Omega_m h^{3.4}$:
\begin{equation}
\frac{\delta H_0}{H_0} = \frac{1}{1.4} \left[ \frac{\delta\left( \Omega_m h^{3.4}\right)}{\Omega_m h^{3.4}}-\frac{\delta\left( \Omega_m h^2\right)}{\Omega_m h^2}\right] \, .
\end{equation}
When analyzing the $\mathsf{Planck_{15}TT}$ dataset, we have found that both $\om$ and $\omm$ shift to larger values in \LCDM$+\geff$, so that it is not evident, a priori, in which direction $H_0$ should move. However, we find that the relative shift in $\om$ is larger, and this results in a negative, albeit small, shift in $H_0$. On the other hand, including also the polarization data, the shift in $\om$ changes sign, as reported above, and both $\om$ and $\omm$ drive $H_0$ to a larger value. For what concerns $\sigma_8$, it is positively correlated to $A_s$, $n_s$ and $\om$. We have verified that the shifts in these parameters roughly compensate, leading for both datasets to a small upward shift in $\sigma_8$.
 
In Fig.~\ref{FIG:tripl} we present the 1D and 2D posteriors for some of the parameters that show the largest shifts between the standard $\Lambda$CDM and the $\Lambda$CDM+$\geff$ model when the $\mathsf{Planck_{15}TTTEEE}$ dataset is used. 
 
\section{Secret neutrino interactions in Majoron models \label{sec:PP}}
The idea at the basis of Majoron models is that lepton number $L$, which is necessarily violated if neutrinos are Majorana particles, is spontaneously broken globally \cite{Chikashige:1980ui,Schechter:1981cv}. The Majoron is the massless Goldstone boson that appears in the theory once the $L$ symmetry is broken. In this framework, a dynamical realization of the see-saw mechanism is achieved, since the vacuum expectation value (vev) $v_\sigma$ of the parent field $\sigma$ of the Majoron generates the ``large'' Majorana term in the neutrino mass matrix. Once the mass matrix is diagonalized, two massive Majorana neutrinos emerge (per each generation), with masses $m_\mathrm{light} \approx v_\Phi^2/v_\sigma$ and $m_\mathrm{heavy} \approx v_\sigma$, where $v_\Phi \ll v_\sigma$ is the vev of the Standard Model Higgs doublet.
Diagonalization of the mass matrix also yields Majoron-neutrino Yukawa interactions, which might be responsible for neutrino-neutrino scatterings
like those considered in this work.

In more detail, in Majoron models $n$ right-handed neutrinos $\nu_R$ are added to the particle content of the SM, together
with a complex singlet\footnote{See Ref.~\cite{Gelmini:1980re} for a model in which the Majoron is instead a Higgs triplet.} Higgs field $\sigma$, the parent field of the Majoron. This alllows
to write the $SU(2)_L\times U(1)$ invariant Lagrangian (for $n=1$):
\begin{equation}
{\mathcal L} = - y_\Phi \bar{L}_L \tilde\Phi \nu_R - y_\sigma \bar \nu^c_R\sigma \nu_R + \mathrm{h.c.}\, ,
\label{eq:Lseesaw}
\end{equation}
where $L_L=(\nu_L,\,\ell^-_L)^T$ is the left-handed lepton doublet, $\Phi=(\Phi^+,\,\Phi^0)^T$ is the standard model Higgs doublet, $\tilde \Phi = i \sigma_2 \Phi^*$, and
$c$ stands for the charge conjugate partner, i.e. $\nu_R^c = \mathcal{C}\bar\nu_R^T$, with $\mathcal{C}$ being the charge conjugation matrix.
In the unitary gauge, we can write the neutrino-Higgs Yukawa lagrangian as
\begin{equation}
{\mathcal L}_{\nu\Phi} = - \frac{y_\Phi}{\sqrt{2}}\left( v_\Phi + H\right) \left( \bar\nu_L \nu_R +\bar\nu_R \nu_L \right) \, ,
\end{equation}
where $v_\Phi$ is the vev of the Higgs doublet. Similarly, writing $\sigma = (v_\sigma + \rho + i J)/\sqrt{2}$, yields the neutrino-singlet Yukawa lagrangian:
\begin{equation}
{\mathcal L}_{\nu\sigma} = - \frac{y_\sigma}{\sqrt{2}} \left( v_\sigma + \rho + i J \right) \bar\nu_R^c \nu_R + \mathrm{h.c.} \, .
\label{eq:Lnus}
\end{equation}
The imaginary part of $\sigma$ is the Majoron $J$.
The part containing the vevs of the scalar fields generates mass terms for neutrinos:
\begin{equation}
{\mathcal L}_\mathrm{mass} = - \frac{y_\Phi v_\Phi}{\sqrt{2}} \left( \bar\nu_L \nu_R +\bar\nu_R \nu_L \right) 
- \frac{y_\sigma v_\sigma}{\sqrt{2}} \left( \bar\nu_R^c \nu_R + \bar \nu_R \nu_R^c \right) \, .
\end{equation}
In particular, this is a Dirac-Majorana mass term with Dirac mass $m_D\equiv y_\Phi v_\Phi/\sqrt{2}$ and
Majorana mass for the right-handend neutrinos $m_R/2 \equiv y_\sigma v_\sigma/\sqrt{2}$.
Diagonalization of the mass matrix in the see-saw limit $m_R \gg m_D$ (i.e., $v_\sigma \gg v_\Phi$) yields two 
eigenstates with definite masses
\begin{align}
&m_1 \simeq \frac{m_D^2}{m_R}\, , \\
&m_2 \simeq m_R + \frac{m_D^2}{m_R} \, ,
\end{align}
with $m_2 \gg m_1$.
The corresponding left-handed  mass eigenstates $\nu_{1L}$ and $\nu_{2L}$, expressed in terms of $\nu_L$ and $\nu_R^c$, are readily found to be
\begin{align}
&\nu_{1L} = - i \nu_L  + i \frac{m_D}{m_R} \nu_R^c \, ,\\
&\nu_{2L} = \nu_R^c  + \frac{m_D}{m_R} \nu_L \, .
\label{eq:eigvec}
\end{align}
From this last expression it is clear that $\nu_L$ and $\nu^c_R$ mix to form a ``light'' and a ``heavy'' eigenstate. The small mixing angle $\theta$ is given by $\tan 2\theta = 2 m_D/m_R \ll 1$. One can build the two Majorana fields $\nu_i \equiv \nu_{iL} + \nu^c_{iL}$ $(i=1,\,2$) that clearly satisfy the condition $\nu_i = \nu_i^c$.

Now let us turn our attention to the part of the lagrangian responsible for the neutrino-Majoron interactions.
From Eq.~\ref{eq:Lnus}, we get:
\begin{align}
{\mathcal L}_{\nu\sigma} \supset {\mathcal L}_{\nu J} = - i \frac{y_\sigma}{\sqrt 2}J\left(\bar\nu_R^c\nu_R -\bar\nu_R\nu_R^c\right) \, .
\end{align}
Inverting Eqs.~(\ref{eq:eigvec}) to express $\nu_R^c$ in terms of the mass eigenstates, and using the Dirac matrix $\gamma_5$ to express $\nu_{1L}$ and $\nu_{2L}$ as the left-handed projections of $\nu_1$ and $\nu_2$ finally yields:
\begin{widetext}
\begin{equation}
{\mathcal L}_{\nu J} = -i \frac{y_\sigma}{\sqrt 2}J \left[
\bar \nu_2 \gamma_5 \nu_2 
- i \frac{m_D}{m_R} \left( \bar \nu_1 \nu_2 +\bar \nu_2 \nu_1\right) 
- \frac{m_D^2}{m_R^2} \, \bar\nu_1\gamma_5\nu_1 
\right]
\label{eq:lagJnu}
\end{equation}
\end{widetext}

If we now concentrate on the last term in Eq. (\ref{eq:lagJnu}), this corresponds to an interaction term of the form\footnote{The $1/2$ in Eq (\ref{eq:lagJnu1})  is introduced in the case of Majorana particles so that the corresponding Feynman rules will weight each $J\nu_1\nu_1$ vertex with $g$ instead than $2g$.} 

\begin{equation}
{\mathcal L} = \frac{ig}{2}  J \, \bar{\nu}_1 \gamma^5 \nu_1 \, ,
\label{eq:lagJnu1}
\end{equation}
coupling a Majoron and two light Majorana neutrinos, with a coupling constant $g~\equiv \sqrt{2} y_\sigma m_D^2m_R^{-2} = m_1/v_\sigma$. The presence of this vertex allows light neutrino scattering $\nu_1\nu_1 \to \nu_1\nu_1$ mediated by a Majoron. The tree-level amplitude for this process can be computed following the Feynman rules for Majorana particles\footnote{Feyman rules for Majorana particles are summarized e.g. in  Refs. \cite{Srednicki:2007qs,Dreiner:2008tw,Valle:2015pba}.}, and yields, in the limit of very small $\nu_1$ and $J$ masses, an unpolarized total cross section:
\begin{equation}
\sigma(\nu_1\nu_1\to \nu_1\nu_1) =\frac{g^4}{4\pi s} \, ,
\end{equation}
where $s$ is the square of the center-of-mass energy. The cross section enters in the Boltzmann evolution of the cosmological neutrino gas in the form of the thermally-averaged cross section times velocity $\langle\sigma v \rangle$ \cite{Kolb:1990vq}; this can be readily computed to give
\begin{equation}
\langle\sigma v \rangle = \frac{\pi^3}{2592\,\zeta(3)^2}\times\frac{g^4}{T^2} \, ,
\end{equation}
which has exactly the same form as Eq.~(\ref{eq:sv}), with $\xi = \pi^3/2592\,\zeta(3)^2 \simeq 8.3\times 10^{-3}$. Here $\zeta(3)$ is the Riemann Zeta function of 3. In order to connect with the results shown in the previous sections, it is enough to note that $g_\mathrm{eff} \equiv \xi^{1/4} g \simeq 0.3 g$ for the class of models considered here. Then, if the neutrino interaction Lagrangian is extended by adding a term of the form (\ref{eq:lagJnu1}), the limits reported in Tabs. \ref{tab:1} and \ref{tab:2} imply:
\begin{equation}
g \lesssim 7 \times 10^{-7} \, \qquad(95\%\,\mathrm{C.I.}). 
\label{eq:gbound}
\end{equation}
If we further assume that the interaction Lagrangian (\ref{eq:lagJnu1}) stems from (\ref{eq:Lseesaw}), i.e. from a dynamical realization of the see-saw mechanism, as illustrated at the beginning of this section, the following bound can be derived from (\ref{eq:gbound}):
\begin{equation}
v_\sigma \gtrsim (1.4\times 10^6) m_\nu, \,
\end{equation}
where we have used $g=m_1/v_\sigma$, and $m_\nu \equiv m_1$.

This picture is still qualitatively valid if one considers its extension to the case of $n>1$ neutrino families. 
In that case, the mass spectrum will still split into $n$ light and $n$ heavy neutrinos. 
A perturbative expansion of the coupling matrix between the Majoron and light neutrinos belonging to mass eigenstates $i$ and $j$ 
yields \cite{Schechter:1981cv} $g_{ij} = (m_i/v_\sigma) \delta_{ij}\sim \epsilon^2 \delta_{ij}$ to leading order in the small
quantity $\epsilon\equiv m_D/m_R$. In other words, at leading order couplings in the mass basis are diagonal, and, as in the $n=1$ case, directly proportional to the neutrino masses. Off-diagonal couplings $g_{ij}$ ($i\neq j$), that might generate neutrino decays, are suppressed like $\epsilon^4$.

In our treatment of the collisional Boltzmann equation, outlined in Sec.~\ref{sec:2}, we have implicitly assumed that 
the couplings are diagonal and do not depend on the particular mass eigenstate, i.e. that the coupling matrix is proportional
to the identity matrix. In the framework discussed in this section, this amounts to the assumption that neutrino masses
are nearly degenerate. However we expect our results to still hold qualitatively even if neutrino masses are hierarchical.

\section{Conclusions}
\label{sec:C}

Cosmological observations are in excellent agreement with the standard picture of neutrino decoupling at $T\simeq 1\,\MeV$ 
and being free-streaming afterwards. However, non-standard neutrino interactions might change this picture, leaving 
a distinct imprint on the CMB anisotropy pattern. We have analysed the most recent publicly available Planck data 
in a framework in which neutrinos have secret self-interactions, with the complementary goals of i) testing the free-streaming
nature of the cosmic neutrinos, and ii) constraining the strength of the self interaction, parametrised in the form of an 
effective coupling constant $\geff$. We have considered the case of a very light mediator $\phi$ for the new interaction, and thus an interaction rate for $\nu\nu \to \nu\nu$ scaling as $\Gamma\propto \geff^4 T$,
corresponding to a thermally-averaged cross section $\langle \sigma v \rangle = \geff^4/T^2 $. In this framework,
cosmic neutrinos, after electroweak decoupling, become collisional again at some later time. This leaves a distinct signature in the CMB power spectrum, in the form of an amplitude boost and a phase shift at the scales that enter the horizon while neutrinos are collisional, as shown in Sec.~\ref{sec:2}. 

Note that it is possible to consider other classes of models than those considered here. For instance, another well-studied 
scenario is the one in which the secret mediator is very heavy, so that the self-interaction reduces to a Fermi-like four-point interaction,
and the scattering rate $\Gamma \propto G_X^2 T^5$ in this case. This has been recently proposed by Kreisch et al. \cite{Kreisch:2019yzn} as an alternative to the standard $\Lambda$CDM model solving the tensions between CMB and low-redshift measurements of $H_0$ and $\sigma_8$. We have instead focused on the complementary limiting case of a very light mediator. Given the results of Ref.~\cite{Kreisch:2019yzn}, where two regions of high probability are found in the parameter space of the model, corresponding to ``moderately'' and ``strongly'' interacting neutrinos respectively, it is worth considering why our analysis finds that the probability distribution is unimodal, with $\geff$ compatible with 0 in most cases. The interest in this behaviour is further justified by the fact that the ``strongly interacting'' region of Kreisch et al. provides a good fit of both high- and low-redshift cosmological data, alleviating the $H_0$ and $\sigma_8$ tension. We argue that the reason for this difference lies on the different scales that are affected by neutrino collisions in the case of a heavy, as opposed to a light, mediator. In the former case, the interaction affects small scales, since neutrinos were collisional at early times; this is one of the key aspects that allows to fit the cosmological data in an enlarged model. In the latter case, instead, larger scales are affected, because neutrinos become collisional again, after electroweak decoupling, at late times. This is evident by comparing the plots of the residuals in Fig~\ref{FIG:pws} of this paper with Fig.~1 of Ref.~\cite{Kreisch:2019yzn}. In addition to this, there are other differences in the two analyses, that might also play a minor role, like the fact that Kreisch et al. consider massive neutrinos: this helps in alleviating the $H_0$ tension. Finally, as already noted, the algorithm that we employ to explore the parameter space is not specifically designed to find maxima of multimodal distributions. We note however that using a logarithmic prior on $\geff$, that should allow to span several orders of magnitude in the parameter, did not show any hint for the presence of a maximum at larger values of $\geff$. In any case we plan to perform a dedicated analysis in a future work.

Throughout our analysis, we have approximated neutrinos as being massless when calculating both the scattering cross sections,
and their effect on the evolution of cosmological perturbations. Although this might seem as a strong assumption at first, note that most of
the constraining power of the CMB to secret interactions comes from redshifts where neutrino masses are negligible. Moreover,
given the sensitivity of current experiments, our assumption basically amounts to fixing the sum of neutrino masses to the minimum
value allowed by oscillation experiments, $\sum m_\nu = 0.06\,\eV$. On the other hand, this approximation is presumably not accurate enough for future data, given the sensitivity of next-generation experiments. 

We have found a 95\% credible interval for the effective coupling constant $\geff^4 < 2.35 \times 10^{-27}$ from a baseline dataset
consisting of the Planck 2015 temperature, large-scale polarization and lensing data. This bound becomes $\geff^4 < 1.64 \times 10^{-27}$
when Planck 2015 small-scale polarization is also included, and represents an improvement of more than a factor 3 with respect to the limits that we previously derived using the Planck 2013 data \cite{Forastieri:2015paa}. We have found no significant improvement in these bounds when also considering geometrical information in the form of measurements of the BAO scale, astrophysical determinations of the Hubble parameter, and observations of type IA supernovae. From the point of view of the cosmological evolution, our results are more easily interpreted by remembering that $ \geff\equiv\langle \sigma v \rangle T^2$. The bounds can also be easily rephrased in terms of the redshift $
\zrec$ at which neutrinos cease to free-stream: $\zrec < 3800$ for the baseline dataset, and $\zrec <2300$ when small-scale polarization is added (95\% credible intervals). Thus we find that neutrinos are free-streaming at least until close to the time of matter-radiation equality. Finally, we find that tensions between CMB and low-redshift measurements of the expansion rate and of the amplitude of matter fluctuations
are not alleviated when NS$\nu$I of this kind are considered. We caution, however, that we have not employed techniques specifically targeted to the search of multiple peaks in the joint posterior distribution of the model parameters.
The results summarized up to this point, reported in Sec.~\ref{sec:R}, do not depend on the underlying particle physics model, as long as the assumption that the 
mediator $\phi$ is very light holds.

On the other hand, from the point of view of particle physics, the quantity $\geff$ can be seen as a proxy for the actual coupling constant $g$ appearing in the interaction Lagrangian of neutrinos, in the sense that $\geff \sim g$ up to a numerical factor, typically of order unity. We can reasonably expect that, barring cancellations, our results give an order-of-magnitude constraint for interaction terms (sketchily) of the form $g\bar\nu\Gamma^{A}\phi_A\nu$, where $A$ is some combination of Lorentz indices and $\Gamma^A$ is some combination of Dirac $\gamma$ matrices, and the mediator $\phi_A$ is very light. The upper limit for $g$ lies somewhere in the ballpark of (few $\times 10^{-7}$), where a more precise value can be obtained once the form of the Lagrangian is specified. Even without picking up a specific model, this constraint is anyway better than those that can be obtained with
other probes (e.g. neutrinoless double $\beta$ decay).
We have explicitly shown in Sec.~\ref{sec:PP} that if we consider a pseudoscalar interaction between Majorana neutrinos, i.e. an interaction Lagrangian of the form ${\mathcal L}_\mathrm{int} = \frac{1}{2} g\,\phi \bar{\nu} (i\gamma ^5) \nu$, our limits on 
$\langle \sigma v \rangle T^2$ imply $g < 7\times 10^{-7}$. By further assuming that the pseudoscalar interaction originates
as a consequence of the (lepton-number-breaking) mechanism that generates small neutrino masses, as in Majoron models, we have been able to put the following loose constraint on the scale $v_\sigma$ of lepton number breaking: $v_\sigma > (1.4 \times 10^{6})\,m_\nu$.

\begin{acknowledgments}
We acknowledge support from the ASI grant 2016-24-H.0 COSMOS ``Attivit\`a di studio per la comunit\`a scientifica di cosmologia'', from INFN through the InDark and Gruppo IV fundings, and from the ASI/INAF Agreement I/072/09/0 for the Planck LFI Activity of Phase E2. We thank Nikita Blinov, Sam McDermott, Martina Gerbino, Pedro Machado, Carlos Wagner, Lian-Tao Wang for useful discussion. 
\end{acknowledgments}

\bibliography{bibliography}

\begin{thebibliography}{70}%
\makeatletter
\providecommand \@ifxundefined [1]{%
 \@ifx{#1\undefined}
}%
\providecommand \@ifnum [1]{%
 \ifnum #1\expandafter \@firstoftwo
 \else \expandafter \@secondoftwo
 \fi
}%
\providecommand \@ifx [1]{%
 \ifx #1\expandafter \@firstoftwo
 \else \expandafter \@secondoftwo
 \fi
}%
\providecommand \natexlab [1]{#1}%
\providecommand \enquote  [1]{``#1''}%
\providecommand \bibnamefont  [1]{#1}%
\providecommand \bibfnamefont [1]{#1}%
\providecommand \citenamefont [1]{#1}%
\providecommand \href@noop [0]{\@secondoftwo}%
\providecommand \href [0]{\begingroup \@sanitize@url \@href}%
\providecommand \@href[1]{\@@startlink{#1}\@@href}%
\providecommand \@@href[1]{\endgroup#1\@@endlink}%
\providecommand \@sanitize@url [0]{\catcode `\\12\catcode `\$12\catcode
  `\&12\catcode `\#12\catcode `\^12\catcode `\_12\catcode `\%12\relax}%
\providecommand \@@startlink[1]{}%
\providecommand \@@endlink[0]{}%
\providecommand \url  [0]{\begingroup\@sanitize@url \@url }%
\providecommand \@url [1]{\endgroup\@href {#1}{\urlprefix }}%
\providecommand \urlprefix  [0]{URL }%
\providecommand \Eprint [0]{\href }%
\providecommand \doibase [0]{http://dx.doi.org/}%
\providecommand \selectlanguage [0]{\@gobble}%
\providecommand \bibinfo  [0]{\@secondoftwo}%
\providecommand \bibfield  [0]{\@secondoftwo}%
\providecommand \translation [1]{[#1]}%
\providecommand \BibitemOpen [0]{}%
\providecommand \bibitemStop [0]{}%
\providecommand \bibitemNoStop [0]{.\EOS\space}%
\providecommand \EOS [0]{\spacefactor3000\relax}%
\providecommand \BibitemShut  [1]{\csname bibitem#1\endcsname}%
\let\auto@bib@innerbib\@empty
\bibitem [{\citenamefont {Bashinsky}\ and\ \citenamefont
  {Seljak}(2004)}]{Bashinsky:2003tk}%
  \BibitemOpen
  \bibfield  {author} {\bibinfo {author} {\bibfnamefont {S.}~\bibnamefont
  {Bashinsky}}\ and\ \bibinfo {author} {\bibfnamefont {U.}~\bibnamefont
  {Seljak}},\ }\href {\doibase 10.1103/PhysRevD.69.083002} {\bibfield
  {journal} {\bibinfo  {journal} {Phys. Rev.}\ }\textbf {\bibinfo {volume}
  {D69}},\ \bibinfo {pages} {083002} (\bibinfo {year} {2004})},\ \Eprint
  {http://arxiv.org/abs/astro-ph/0310198} {arXiv:astro-ph/0310198 [astro-ph]}
  \BibitemShut {NoStop}%
\bibitem [{\citenamefont {Follin}\ \emph {et~al.}(2015)\citenamefont {Follin},
  \citenamefont {Knox}, \citenamefont {Millea},\ and\ \citenamefont
  {Pan}}]{Follin:2015hya}%
  \BibitemOpen
  \bibfield  {author} {\bibinfo {author} {\bibfnamefont {B.}~\bibnamefont
  {Follin}}, \bibinfo {author} {\bibfnamefont {L.}~\bibnamefont {Knox}},
  \bibinfo {author} {\bibfnamefont {M.}~\bibnamefont {Millea}}, \ and\ \bibinfo
  {author} {\bibfnamefont {Z.}~\bibnamefont {Pan}},\ }\href {\doibase
  10.1103/PhysRevLett.115.091301} {\bibfield  {journal} {\bibinfo  {journal}
  {Phys. Rev. Lett.}\ }\textbf {\bibinfo {volume} {115}},\ \bibinfo {pages}
  {091301} (\bibinfo {year} {2015})},\ \Eprint
  {http://arxiv.org/abs/1503.07863} {arXiv:1503.07863 [astro-ph.CO]}
  \BibitemShut {NoStop}%
\bibitem [{\citenamefont {Baumann}\ \emph {et~al.}(2016)\citenamefont
  {Baumann}, \citenamefont {Green}, \citenamefont {Meyers},\ and\ \citenamefont
  {Wallisch}}]{Baumann:2015rya}%
  \BibitemOpen
  \bibfield  {author} {\bibinfo {author} {\bibfnamefont {D.}~\bibnamefont
  {Baumann}}, \bibinfo {author} {\bibfnamefont {D.}~\bibnamefont {Green}},
  \bibinfo {author} {\bibfnamefont {J.}~\bibnamefont {Meyers}}, \ and\ \bibinfo
  {author} {\bibfnamefont {B.}~\bibnamefont {Wallisch}},\ }\href {\doibase
  10.1088/1475-7516/2016/01/007} {\bibfield  {journal} {\bibinfo  {journal}
  {JCAP}\ }\textbf {\bibinfo {volume} {1601}},\ \bibinfo {pages} {007}
  (\bibinfo {year} {2016})},\ \Eprint {http://arxiv.org/abs/1508.06342}
  {arXiv:1508.06342 [astro-ph.CO]} \BibitemShut {NoStop}%
\bibitem [{\citenamefont {Chikashige}\ \emph {et~al.}(1981)\citenamefont
  {Chikashige}, \citenamefont {Mohapatra},\ and\ \citenamefont
  {Peccei}}]{Chikashige:1980ui}%
  \BibitemOpen
  \bibfield  {author} {\bibinfo {author} {\bibfnamefont {Y.}~\bibnamefont
  {Chikashige}}, \bibinfo {author} {\bibfnamefont {R.~N.}\ \bibnamefont
  {Mohapatra}}, \ and\ \bibinfo {author} {\bibfnamefont {R.~D.}\ \bibnamefont
  {Peccei}},\ }\href {\doibase 10.1016/0370-2693(81)90011-3} {\bibfield
  {journal} {\bibinfo  {journal} {Phys. Lett.}\ }\textbf {\bibinfo {volume}
  {98B}},\ \bibinfo {pages} {265} (\bibinfo {year} {1981})}\BibitemShut
  {NoStop}%
\bibitem [{\citenamefont {Gelmini}\ and\ \citenamefont
  {Roncadelli}(1981)}]{Gelmini:1980re}%
  \BibitemOpen
  \bibfield  {author} {\bibinfo {author} {\bibfnamefont {G.~B.}\ \bibnamefont
  {Gelmini}}\ and\ \bibinfo {author} {\bibfnamefont {M.}~\bibnamefont
  {Roncadelli}},\ }\href {\doibase 10.1016/0370-2693(81)90559-1} {\bibfield
  {journal} {\bibinfo  {journal} {Phys. Lett.}\ }\textbf {\bibinfo {volume}
  {99B}},\ \bibinfo {pages} {411} (\bibinfo {year} {1981})}\BibitemShut
  {NoStop}%
\bibitem [{\citenamefont {Schechter}\ and\ \citenamefont
  {Valle}(1982)}]{Schechter:1981cv}%
  \BibitemOpen
  \bibfield  {author} {\bibinfo {author} {\bibfnamefont {J.}~\bibnamefont
  {Schechter}}\ and\ \bibinfo {author} {\bibfnamefont {J.~W.~F.}\ \bibnamefont
  {Valle}},\ }\href {\doibase 10.1103/PhysRevD.25.774} {\bibfield  {journal}
  {\bibinfo  {journal} {Phys. Rev.}\ }\textbf {\bibinfo {volume} {D25}},\
  \bibinfo {pages} {774} (\bibinfo {year} {1982})}\BibitemShut {NoStop}%
\bibitem [{\citenamefont {Valle}\ and\ \citenamefont
  {Romao}(2015)}]{Valle:2015pba}%
  \BibitemOpen
  \bibfield  {author} {\bibinfo {author} {\bibfnamefont {J.~W.~F.}\
  \bibnamefont {Valle}}\ and\ \bibinfo {author} {\bibfnamefont {J.~C.}\
  \bibnamefont {Romao}},\ }\href
  {http://eu.wiley.com/WileyCDA/WileyTitle/productCd-3527411976.html} {\emph
  {\bibinfo {title} {{Neutrinos in high energy and astroparticle physics}}}},\
  Physics textbook\ (\bibinfo  {publisher} {Wiley-VCH},\ \bibinfo {address}
  {Weinheim},\ \bibinfo {year} {2015})\BibitemShut {NoStop}%
\bibitem [{\citenamefont {Bardin}\ \emph {et~al.}(1970)\citenamefont {Bardin},
  \citenamefont {Bilenky},\ and\ \citenamefont {Pontecorvo}}]{bardin1970nu}%
  \BibitemOpen
  \bibfield  {author} {\bibinfo {author} {\bibfnamefont {D.~Y.}\ \bibnamefont
  {Bardin}}, \bibinfo {author} {\bibfnamefont {S.}~\bibnamefont {Bilenky}}, \
  and\ \bibinfo {author} {\bibfnamefont {B.}~\bibnamefont {Pontecorvo}},\
  }\href@noop {} {\bibfield  {journal} {\bibinfo  {journal} {Physics Letters
  B}\ }\textbf {\bibinfo {volume} {32}},\ \bibinfo {pages} {121} (\bibinfo
  {year} {1970})}\BibitemShut {NoStop}%
\bibitem [{\citenamefont {Bilenky}\ \emph {et~al.}(1993)\citenamefont
  {Bilenky}, \citenamefont {Bilenky},\ and\ \citenamefont
  {Santamaria}}]{bilenky1993invisible}%
  \BibitemOpen
  \bibfield  {author} {\bibinfo {author} {\bibfnamefont {M.}~\bibnamefont
  {Bilenky}}, \bibinfo {author} {\bibfnamefont {S.~M.}\ \bibnamefont
  {Bilenky}}, \ and\ \bibinfo {author} {\bibfnamefont {A.}~\bibnamefont
  {Santamaria}},\ }\href@noop {} {\bibfield  {journal} {\bibinfo  {journal}
  {Physics Letters B}\ }\textbf {\bibinfo {volume} {301}},\ \bibinfo {pages}
  {287} (\bibinfo {year} {1993})}\BibitemShut {NoStop}%
\bibitem [{\citenamefont {Bilenky}\ and\ \citenamefont
  {Santamaria}(1999)}]{bilenky1999secret}%
  \BibitemOpen
  \bibfield  {author} {\bibinfo {author} {\bibfnamefont {M.}~\bibnamefont
  {Bilenky}}\ and\ \bibinfo {author} {\bibfnamefont {A.}~\bibnamefont
  {Santamaria}},\ }\href@noop {} {\bibfield  {journal} {\bibinfo  {journal}
  {arXiv preprint hep-ph/9908272}\ } (\bibinfo {year} {1999})}\BibitemShut
  {NoStop}%
\bibitem [{\citenamefont {Akimov}\ \emph {et~al.}(2017)\citenamefont {Akimov}
  \emph {et~al.}}]{Akimov:2017ade}%
  \BibitemOpen
  \bibfield  {author} {\bibinfo {author} {\bibfnamefont {D.}~\bibnamefont
  {Akimov}} \emph {et~al.} (\bibinfo {collaboration} {COHERENT}),\ }\href
  {\doibase 10.1126/science.aao0990} {\bibfield  {journal} {\bibinfo  {journal}
  {Science}\ }\textbf {\bibinfo {volume} {357}},\ \bibinfo {pages} {1123}
  (\bibinfo {year} {2017})},\ \Eprint {http://arxiv.org/abs/1708.01294}
  {arXiv:1708.01294 [nucl-ex]} \BibitemShut {NoStop}%
\bibitem [{\citenamefont {Barranco}\ \emph {et~al.}(2007)\citenamefont
  {Barranco}, \citenamefont {Miranda},\ and\ \citenamefont
  {Rashba}}]{Barranco:2007tz}%
  \BibitemOpen
  \bibfield  {author} {\bibinfo {author} {\bibfnamefont {J.}~\bibnamefont
  {Barranco}}, \bibinfo {author} {\bibfnamefont {O.~G.}\ \bibnamefont
  {Miranda}}, \ and\ \bibinfo {author} {\bibfnamefont {T.~I.}\ \bibnamefont
  {Rashba}},\ }\href {\doibase 10.1103/PhysRevD.76.073008} {\bibfield
  {journal} {\bibinfo  {journal} {Phys. Rev.}\ }\textbf {\bibinfo {volume}
  {D76}},\ \bibinfo {pages} {073008} (\bibinfo {year} {2007})},\ \Eprint
  {http://arxiv.org/abs/hep-ph/0702175} {arXiv:hep-ph/0702175 [hep-ph]}
  \BibitemShut {NoStop}%
\bibitem [{\citenamefont {deNiverville}\ \emph {et~al.}(2015)\citenamefont
  {deNiverville}, \citenamefont {Pospelov},\ and\ \citenamefont
  {Ritz}}]{deNiverville:2015mwa}%
  \BibitemOpen
  \bibfield  {author} {\bibinfo {author} {\bibfnamefont {P.}~\bibnamefont
  {deNiverville}}, \bibinfo {author} {\bibfnamefont {M.}~\bibnamefont
  {Pospelov}}, \ and\ \bibinfo {author} {\bibfnamefont {A.}~\bibnamefont
  {Ritz}},\ }\href {\doibase 10.1103/PhysRevD.92.095005} {\bibfield  {journal}
  {\bibinfo  {journal} {Phys. Rev.}\ }\textbf {\bibinfo {volume} {D92}},\
  \bibinfo {pages} {095005} (\bibinfo {year} {2015})},\ \Eprint
  {http://arxiv.org/abs/1505.07805} {arXiv:1505.07805 [hep-ph]} \BibitemShut
  {NoStop}%
\bibitem [{\citenamefont {Dutta}\ \emph {et~al.}(2016)\citenamefont {Dutta},
  \citenamefont {Mahapatra}, \citenamefont {Strigari},\ and\ \citenamefont
  {Walker}}]{Dutta:2015vwa}%
  \BibitemOpen
  \bibfield  {author} {\bibinfo {author} {\bibfnamefont {B.}~\bibnamefont
  {Dutta}}, \bibinfo {author} {\bibfnamefont {R.}~\bibnamefont {Mahapatra}},
  \bibinfo {author} {\bibfnamefont {L.~E.}\ \bibnamefont {Strigari}}, \ and\
  \bibinfo {author} {\bibfnamefont {J.~W.}\ \bibnamefont {Walker}},\ }\href
  {\doibase 10.1103/PhysRevD.93.013015} {\bibfield  {journal} {\bibinfo
  {journal} {Phys. Rev.}\ }\textbf {\bibinfo {volume} {D93}},\ \bibinfo {pages}
  {013015} (\bibinfo {year} {2016})},\ \Eprint
  {http://arxiv.org/abs/1508.07981} {arXiv:1508.07981 [hep-ph]} \BibitemShut
  {NoStop}%
\bibitem [{\citenamefont {Liao}\ and\ \citenamefont
  {Marfatia}(2017)}]{Liao:2017uzy}%
  \BibitemOpen
  \bibfield  {author} {\bibinfo {author} {\bibfnamefont {J.}~\bibnamefont
  {Liao}}\ and\ \bibinfo {author} {\bibfnamefont {D.}~\bibnamefont
  {Marfatia}},\ }\href {\doibase 10.1016/j.physletb.2017.10.046} {\bibfield
  {journal} {\bibinfo  {journal} {Phys. Lett.}\ }\textbf {\bibinfo {volume}
  {B775}},\ \bibinfo {pages} {54} (\bibinfo {year} {2017})},\ \Eprint
  {http://arxiv.org/abs/1708.04255} {arXiv:1708.04255 [hep-ph]} \BibitemShut
  {NoStop}%
\bibitem [{\citenamefont {Farzan}\ \emph {et~al.}(2018)\citenamefont {Farzan},
  \citenamefont {Lindner}, \citenamefont {Rodejohann},\ and\ \citenamefont
  {Xu}}]{Farzan:2018gtr}%
  \BibitemOpen
  \bibfield  {author} {\bibinfo {author} {\bibfnamefont {Y.}~\bibnamefont
  {Farzan}}, \bibinfo {author} {\bibfnamefont {M.}~\bibnamefont {Lindner}},
  \bibinfo {author} {\bibfnamefont {W.}~\bibnamefont {Rodejohann}}, \ and\
  \bibinfo {author} {\bibfnamefont {X.-J.}\ \bibnamefont {Xu}},\ }\href
  {\doibase 10.1007/JHEP05(2018)066} {\bibfield  {journal} {\bibinfo  {journal}
  {JHEP}\ }\textbf {\bibinfo {volume} {05}},\ \bibinfo {pages} {066} (\bibinfo
  {year} {2018})},\ \Eprint {http://arxiv.org/abs/1802.05171} {arXiv:1802.05171
  [hep-ph]} \BibitemShut {NoStop}%
\bibitem [{\citenamefont {Kolb}\ and\ \citenamefont
  {Turner}(1987)}]{kolb1987supernova}%
  \BibitemOpen
  \bibfield  {author} {\bibinfo {author} {\bibfnamefont {E.~W.}\ \bibnamefont
  {Kolb}}\ and\ \bibinfo {author} {\bibfnamefont {M.~S.}\ \bibnamefont
  {Turner}},\ }\href@noop {} {\bibfield  {journal} {\bibinfo  {journal}
  {Physical Review D}\ }\textbf {\bibinfo {volume} {36}},\ \bibinfo {pages}
  {2895} (\bibinfo {year} {1987})}\BibitemShut {NoStop}%
\bibitem [{\citenamefont {Manohar}(1987)}]{manohar1987limit}%
  \BibitemOpen
  \bibfield  {author} {\bibinfo {author} {\bibfnamefont {A.}~\bibnamefont
  {Manohar}},\ }\href@noop {} {\bibfield  {journal} {\bibinfo  {journal}
  {Physics Letters B}\ }\textbf {\bibinfo {volume} {192}},\ \bibinfo {pages}
  {217} (\bibinfo {year} {1987})}\BibitemShut {NoStop}%
\bibitem [{\citenamefont {Dicus}\ \emph {et~al.}(1989)\citenamefont {Dicus},
  \citenamefont {Nussinov}, \citenamefont {Pal},\ and\ \citenamefont
  {Teplitz}}]{dicus1989implications}%
  \BibitemOpen
  \bibfield  {author} {\bibinfo {author} {\bibfnamefont {D.~A.}\ \bibnamefont
  {Dicus}}, \bibinfo {author} {\bibfnamefont {S.}~\bibnamefont {Nussinov}},
  \bibinfo {author} {\bibfnamefont {P.~B.}\ \bibnamefont {Pal}}, \ and\
  \bibinfo {author} {\bibfnamefont {V.~L.}\ \bibnamefont {Teplitz}},\
  }\href@noop {} {\bibfield  {journal} {\bibinfo  {journal} {Physics Letters
  B}\ }\textbf {\bibinfo {volume} {218}},\ \bibinfo {pages} {84} (\bibinfo
  {year} {1989})}\BibitemShut {NoStop}%
\bibitem [{\citenamefont {Kachelriess}\ \emph {et~al.}(2000)\citenamefont
  {Kachelriess}, \citenamefont {Tom{\`a}s},\ and\ \citenamefont
  {Valle}}]{kachelriess2000supernova}%
  \BibitemOpen
  \bibfield  {author} {\bibinfo {author} {\bibfnamefont {M.}~\bibnamefont
  {Kachelriess}}, \bibinfo {author} {\bibfnamefont {R.}~\bibnamefont
  {Tom{\`a}s}}, \ and\ \bibinfo {author} {\bibfnamefont {J.}~\bibnamefont
  {Valle}},\ }\href@noop {} {\bibfield  {journal} {\bibinfo  {journal}
  {Physical Review D}\ }\textbf {\bibinfo {volume} {62}},\ \bibinfo {pages}
  {023004} (\bibinfo {year} {2000})}\BibitemShut {NoStop}%
\bibitem [{\citenamefont {Zhou}(2011)}]{zhou2011comment}%
  \BibitemOpen
  \bibfield  {author} {\bibinfo {author} {\bibfnamefont {S.}~\bibnamefont
  {Zhou}},\ }\href@noop {} {\bibfield  {journal} {\bibinfo  {journal} {Physical
  Review D}\ }\textbf {\bibinfo {volume} {84}},\ \bibinfo {pages} {038701}
  (\bibinfo {year} {2011})}\BibitemShut {NoStop}%
\bibitem [{\citenamefont {Jeong}\ \emph {et~al.}(2018)\citenamefont {Jeong},
  \citenamefont {Palomares-Ruiz}, \citenamefont {Reno},\ and\ \citenamefont
  {Sarcevic}}]{jeong2018probing}%
  \BibitemOpen
  \bibfield  {author} {\bibinfo {author} {\bibfnamefont {Y.~S.}\ \bibnamefont
  {Jeong}}, \bibinfo {author} {\bibfnamefont {S.}~\bibnamefont
  {Palomares-Ruiz}}, \bibinfo {author} {\bibfnamefont {M.~H.}\ \bibnamefont
  {Reno}}, \ and\ \bibinfo {author} {\bibfnamefont {I.}~\bibnamefont
  {Sarcevic}},\ }\href@noop {} {\bibfield  {journal} {\bibinfo  {journal}
  {Journal of Cosmology and Astroparticle Physics}\ }\textbf {\bibinfo {volume}
  {2018}},\ \bibinfo {pages} {019} (\bibinfo {year} {2018})}\BibitemShut
  {NoStop}%
\bibitem [{\citenamefont {Cremonesi}\ and\ \citenamefont
  {Pavan}(2014)}]{cremonesi2014challenges}%
  \BibitemOpen
  \bibfield  {author} {\bibinfo {author} {\bibfnamefont {O.}~\bibnamefont
  {Cremonesi}}\ and\ \bibinfo {author} {\bibfnamefont {M.}~\bibnamefont
  {Pavan}},\ }\href@noop {} {\bibfield  {journal} {\bibinfo  {journal}
  {Advances in High Energy Physics}\ }\textbf {\bibinfo {volume} {2014}}
  (\bibinfo {year} {2014})}\BibitemShut {NoStop}%
\bibitem [{\citenamefont {Dell'Oro}\ \emph {et~al.}(2016)\citenamefont
  {Dell'Oro}, \citenamefont {Marcocci}, \citenamefont {Viel},\ and\
  \citenamefont {Vissani}}]{dell2016neutrinoless}%
  \BibitemOpen
  \bibfield  {author} {\bibinfo {author} {\bibfnamefont {S.}~\bibnamefont
  {Dell'Oro}}, \bibinfo {author} {\bibfnamefont {S.}~\bibnamefont {Marcocci}},
  \bibinfo {author} {\bibfnamefont {M.}~\bibnamefont {Viel}}, \ and\ \bibinfo
  {author} {\bibfnamefont {F.}~\bibnamefont {Vissani}},\ }\href@noop {}
  {\bibfield  {journal} {\bibinfo  {journal} {Advances in High Energy Physics}\
  }\textbf {\bibinfo {volume} {2016}} (\bibinfo {year} {2016})}\BibitemShut
  {NoStop}%
\bibitem [{\citenamefont {Bilenky}\ and\ \citenamefont
  {Giunti}(2012)}]{bilenky2012neutrinoless}%
  \BibitemOpen
  \bibfield  {author} {\bibinfo {author} {\bibfnamefont {S.}~\bibnamefont
  {Bilenky}}\ and\ \bibinfo {author} {\bibfnamefont {C.}~\bibnamefont
  {Giunti}},\ }\href@noop {} {\bibfield  {journal} {\bibinfo  {journal} {Modern
  Physics Letters A}\ }\textbf {\bibinfo {volume} {27}},\ \bibinfo {pages}
  {1230015} (\bibinfo {year} {2012})}\BibitemShut {NoStop}%
\bibitem [{\citenamefont {Ioka}\ and\ \citenamefont
  {Murase}(2014)}]{Ioka:2014kca}%
  \BibitemOpen
  \bibfield  {author} {\bibinfo {author} {\bibfnamefont {K.}~\bibnamefont
  {Ioka}}\ and\ \bibinfo {author} {\bibfnamefont {K.}~\bibnamefont {Murase}},\
  }\href {\doibase 10.1093/ptep/ptu090} {\bibfield  {journal} {\bibinfo
  {journal} {PTEP}\ }\textbf {\bibinfo {volume} {2014}},\ \bibinfo {pages}
  {061E01} (\bibinfo {year} {2014})},\ \Eprint {http://arxiv.org/abs/1404.2279}
  {arXiv:1404.2279 [astro-ph.HE]} \BibitemShut {NoStop}%
\bibitem [{\citenamefont {Ng}\ and\ \citenamefont {Beacom}(2014)}]{Ng:2014pca}%
  \BibitemOpen
  \bibfield  {author} {\bibinfo {author} {\bibfnamefont {K.~C.~Y.}\
  \bibnamefont {Ng}}\ and\ \bibinfo {author} {\bibfnamefont {J.~F.}\
  \bibnamefont {Beacom}},\ }\href {\doibase 10.1103/PhysRevD.90.065035,
  10.1103/PhysRevD.90.089904} {\bibfield  {journal} {\bibinfo  {journal} {Phys.
  Rev.}\ }\textbf {\bibinfo {volume} {D90}},\ \bibinfo {pages} {065035}
  (\bibinfo {year} {2014})},\ \bibinfo {note} {[Erratum: Phys.
  Rev.D90,no.8,089904(2014)]},\ \Eprint {http://arxiv.org/abs/1404.2288}
  {arXiv:1404.2288 [astro-ph.HE]} \BibitemShut {NoStop}%
\bibitem [{\citenamefont {Blum}\ \emph {et~al.}(2014)\citenamefont {Blum},
  \citenamefont {Hook},\ and\ \citenamefont {Murase}}]{Blum:2014ewa}%
  \BibitemOpen
  \bibfield  {author} {\bibinfo {author} {\bibfnamefont {K.}~\bibnamefont
  {Blum}}, \bibinfo {author} {\bibfnamefont {A.}~\bibnamefont {Hook}}, \ and\
  \bibinfo {author} {\bibfnamefont {K.}~\bibnamefont {Murase}},\ }\href@noop {}
  {\  (\bibinfo {year} {2014})},\ \Eprint {http://arxiv.org/abs/1408.3799}
  {arXiv:1408.3799 [hep-ph]} \BibitemShut {NoStop}%
\bibitem [{\citenamefont {Cherry}\ \emph {et~al.}(2016)\citenamefont {Cherry},
  \citenamefont {Friedland},\ and\ \citenamefont
  {Shoemaker}}]{cherry2016short}%
  \BibitemOpen
  \bibfield  {author} {\bibinfo {author} {\bibfnamefont {J.~F.}\ \bibnamefont
  {Cherry}}, \bibinfo {author} {\bibfnamefont {A.}~\bibnamefont {Friedland}}, \
  and\ \bibinfo {author} {\bibfnamefont {I.~M.}\ \bibnamefont {Shoemaker}},\
  }\href@noop {} {\bibfield  {journal} {\bibinfo  {journal} {arXiv preprint
  arXiv:1605.06506}\ } (\bibinfo {year} {2016})}\BibitemShut {NoStop}%
\bibitem [{\citenamefont {Denton}\ and\ \citenamefont
  {Tamborra}(2018)}]{Denton:2018aml}%
  \BibitemOpen
  \bibfield  {author} {\bibinfo {author} {\bibfnamefont {P.~B.}\ \bibnamefont
  {Denton}}\ and\ \bibinfo {author} {\bibfnamefont {I.}~\bibnamefont
  {Tamborra}},\ }\href {\doibase 10.1103/PhysRevLett.121.121802} {\bibfield
  {journal} {\bibinfo  {journal} {Phys. Rev. Lett.}\ }\textbf {\bibinfo
  {volume} {121}},\ \bibinfo {pages} {121802} (\bibinfo {year} {2018})},\
  \Eprint {http://arxiv.org/abs/1805.05950} {arXiv:1805.05950 [hep-ph]}
  \BibitemShut {NoStop}%
\bibitem [{\citenamefont {Ahlgren}\ \emph {et~al.}(2013)\citenamefont
  {Ahlgren}, \citenamefont {Ohlsson},\ and\ \citenamefont
  {Zhou}}]{ahlgren2013comment}%
  \BibitemOpen
  \bibfield  {author} {\bibinfo {author} {\bibfnamefont {B.}~\bibnamefont
  {Ahlgren}}, \bibinfo {author} {\bibfnamefont {T.}~\bibnamefont {Ohlsson}}, \
  and\ \bibinfo {author} {\bibfnamefont {S.}~\bibnamefont {Zhou}},\ }\href@noop
  {} {\bibfield  {journal} {\bibinfo  {journal} {Physical review letters}\
  }\textbf {\bibinfo {volume} {111}},\ \bibinfo {pages} {199001} (\bibinfo
  {year} {2013})}\BibitemShut {NoStop}%
\bibitem [{\citenamefont {Huang}\ \emph {et~al.}(2018)\citenamefont {Huang},
  \citenamefont {Ohlsson},\ and\ \citenamefont
  {Zhou}}]{huang2018observational}%
  \BibitemOpen
  \bibfield  {author} {\bibinfo {author} {\bibfnamefont {G.-y.}\ \bibnamefont
  {Huang}}, \bibinfo {author} {\bibfnamefont {T.}~\bibnamefont {Ohlsson}}, \
  and\ \bibinfo {author} {\bibfnamefont {S.}~\bibnamefont {Zhou}},\ }\href@noop
  {} {\bibfield  {journal} {\bibinfo  {journal} {Physical Review D}\ }\textbf
  {\bibinfo {volume} {97}},\ \bibinfo {pages} {075009} (\bibinfo {year}
  {2018})}\BibitemShut {NoStop}%
\bibitem [{\citenamefont {Beacom}\ \emph {et~al.}(2004)\citenamefont {Beacom},
  \citenamefont {Bell},\ and\ \citenamefont {Dodelson}}]{Beacom:2004yd}%
  \BibitemOpen
  \bibfield  {author} {\bibinfo {author} {\bibfnamefont {J.~F.}\ \bibnamefont
  {Beacom}}, \bibinfo {author} {\bibfnamefont {N.~F.}\ \bibnamefont {Bell}}, \
  and\ \bibinfo {author} {\bibfnamefont {S.}~\bibnamefont {Dodelson}},\ }\href
  {\doibase 10.1103/PhysRevLett.93.121302} {\bibfield  {journal} {\bibinfo
  {journal} {Phys. Rev. Lett.}\ }\textbf {\bibinfo {volume} {93}},\ \bibinfo
  {pages} {121302} (\bibinfo {year} {2004})},\ \Eprint
  {http://arxiv.org/abs/astro-ph/0404585} {arXiv:astro-ph/0404585 [astro-ph]}
  \BibitemShut {NoStop}%
\bibitem [{\citenamefont {Bell}\ \emph {et~al.}(2006)\citenamefont {Bell},
  \citenamefont {Pierpaoli},\ and\ \citenamefont
  {Sigurdson}}]{bell2006cosmological}%
  \BibitemOpen
  \bibfield  {author} {\bibinfo {author} {\bibfnamefont {N.~F.}\ \bibnamefont
  {Bell}}, \bibinfo {author} {\bibfnamefont {E.}~\bibnamefont {Pierpaoli}}, \
  and\ \bibinfo {author} {\bibfnamefont {K.}~\bibnamefont {Sigurdson}},\
  }\href@noop {} {\bibfield  {journal} {\bibinfo  {journal} {Physical Review
  D}\ }\textbf {\bibinfo {volume} {73}},\ \bibinfo {pages} {063523} (\bibinfo
  {year} {2006})}\BibitemShut {NoStop}%
\bibitem [{\citenamefont {Friedland}\ \emph {et~al.}(2007)\citenamefont
  {Friedland}, \citenamefont {Zurek},\ and\ \citenamefont
  {Bashinsky}}]{Friedland:2007vv}%
  \BibitemOpen
  \bibfield  {author} {\bibinfo {author} {\bibfnamefont {A.}~\bibnamefont
  {Friedland}}, \bibinfo {author} {\bibfnamefont {K.~M.}\ \bibnamefont
  {Zurek}}, \ and\ \bibinfo {author} {\bibfnamefont {S.}~\bibnamefont
  {Bashinsky}},\ }\href@noop {} {\  (\bibinfo {year} {2007})},\ \Eprint
  {http://arxiv.org/abs/0704.3271} {arXiv:0704.3271 [astro-ph]} \BibitemShut
  {NoStop}%
\bibitem [{\citenamefont {Basb{\o}ll}\ \emph {et~al.}(2009)\citenamefont
  {Basb{\o}ll}, \citenamefont {Bjaelde}, \citenamefont {Hannestad},\ and\
  \citenamefont {Raffelt}}]{basboll2009cosmological}%
  \BibitemOpen
  \bibfield  {author} {\bibinfo {author} {\bibfnamefont {A.}~\bibnamefont
  {Basb{\o}ll}}, \bibinfo {author} {\bibfnamefont {O.~E.}\ \bibnamefont
  {Bjaelde}}, \bibinfo {author} {\bibfnamefont {S.}~\bibnamefont {Hannestad}},
  \ and\ \bibinfo {author} {\bibfnamefont {G.~G.}\ \bibnamefont {Raffelt}},\
  }\href@noop {} {\bibfield  {journal} {\bibinfo  {journal} {Physical Review
  D}\ }\textbf {\bibinfo {volume} {79}},\ \bibinfo {pages} {043512} (\bibinfo
  {year} {2009})}\BibitemShut {NoStop}%
\bibitem [{\citenamefont {Cyr-Racine}\ and\ \citenamefont
  {Sigurdson}(2014)}]{Cyr-Racine:2013jua}%
  \BibitemOpen
  \bibfield  {author} {\bibinfo {author} {\bibfnamefont {F.-Y.}\ \bibnamefont
  {Cyr-Racine}}\ and\ \bibinfo {author} {\bibfnamefont {K.}~\bibnamefont
  {Sigurdson}},\ }\href {\doibase 10.1103/PhysRevD.90.123533} {\bibfield
  {journal} {\bibinfo  {journal} {Phys. Rev.}\ }\textbf {\bibinfo {volume}
  {D90}},\ \bibinfo {pages} {123533} (\bibinfo {year} {2014})},\ \Eprint
  {http://arxiv.org/abs/1306.1536} {arXiv:1306.1536 [astro-ph.CO]} \BibitemShut
  {NoStop}%
\bibitem [{\citenamefont {Archidiacono}\ and\ \citenamefont
  {Hannestad}(2014{\natexlab{a}})}]{archidiacono2014updated}%
  \BibitemOpen
  \bibfield  {author} {\bibinfo {author} {\bibfnamefont {M.}~\bibnamefont
  {Archidiacono}}\ and\ \bibinfo {author} {\bibfnamefont {S.}~\bibnamefont
  {Hannestad}},\ }\href@noop {} {\bibfield  {journal} {\bibinfo  {journal}
  {Journal of Cosmology and Astroparticle Physics}\ }\textbf {\bibinfo {volume}
  {2014}},\ \bibinfo {pages} {046} (\bibinfo {year}
  {2014}{\natexlab{a}})}\BibitemShut {NoStop}%
\bibitem [{\citenamefont {Oldengott}\ \emph
  {et~al.}(2015{\natexlab{a}})\citenamefont {Oldengott}, \citenamefont
  {Rampf},\ and\ \citenamefont {Wong}}]{oldengott2015boltzmann}%
  \BibitemOpen
  \bibfield  {author} {\bibinfo {author} {\bibfnamefont {I.~M.}\ \bibnamefont
  {Oldengott}}, \bibinfo {author} {\bibfnamefont {C.}~\bibnamefont {Rampf}}, \
  and\ \bibinfo {author} {\bibfnamefont {Y.~Y.}\ \bibnamefont {Wong}},\
  }\href@noop {} {\bibfield  {journal} {\bibinfo  {journal} {Journal of
  Cosmology and Astroparticle Physics}\ }\textbf {\bibinfo {volume} {2015}},\
  \bibinfo {pages} {016} (\bibinfo {year} {2015}{\natexlab{a}})}\BibitemShut
  {NoStop}%
\bibitem [{\citenamefont {Forastieri}\ \emph {et~al.}(2015)\citenamefont
  {Forastieri}, \citenamefont {Lattanzi},\ and\ \citenamefont
  {Natoli}}]{Forastieri:2015paa}%
  \BibitemOpen
  \bibfield  {author} {\bibinfo {author} {\bibfnamefont {F.}~\bibnamefont
  {Forastieri}}, \bibinfo {author} {\bibfnamefont {M.}~\bibnamefont
  {Lattanzi}}, \ and\ \bibinfo {author} {\bibfnamefont {P.}~\bibnamefont
  {Natoli}},\ }\href {\doibase 10.1088/1475-7516/2015/07/014} {\bibfield
  {journal} {\bibinfo  {journal} {JCAP}\ }\textbf {\bibinfo {volume} {1507}},\
  \bibinfo {pages} {014} (\bibinfo {year} {2015})},\ \Eprint
  {http://arxiv.org/abs/1504.04999} {arXiv:1504.04999 [astro-ph.CO]}
  \BibitemShut {NoStop}%
\bibitem [{\citenamefont {Forastieri}\ \emph {et~al.}(2017)\citenamefont
  {Forastieri}, \citenamefont {Lattanzi}, \citenamefont {Mangano},
  \citenamefont {Mirizzi}, \citenamefont {Natoli},\ and\ \citenamefont
  {Saviano}}]{Forastieri:2017oma}%
  \BibitemOpen
  \bibfield  {author} {\bibinfo {author} {\bibfnamefont {F.}~\bibnamefont
  {Forastieri}}, \bibinfo {author} {\bibfnamefont {M.}~\bibnamefont
  {Lattanzi}}, \bibinfo {author} {\bibfnamefont {G.}~\bibnamefont {Mangano}},
  \bibinfo {author} {\bibfnamefont {A.}~\bibnamefont {Mirizzi}}, \bibinfo
  {author} {\bibfnamefont {P.}~\bibnamefont {Natoli}}, \ and\ \bibinfo {author}
  {\bibfnamefont {N.}~\bibnamefont {Saviano}},\ }\href {\doibase
  10.1088/1475-7516/2017/07/038} {\bibfield  {journal} {\bibinfo  {journal}
  {JCAP}\ }\textbf {\bibinfo {volume} {1707}},\ \bibinfo {pages} {038}
  (\bibinfo {year} {2017})},\ \Eprint {http://arxiv.org/abs/1704.00626}
  {arXiv:1704.00626 [astro-ph.CO]} \BibitemShut {NoStop}%
\bibitem [{\citenamefont {Oldengott}\ \emph {et~al.}(2017)\citenamefont
  {Oldengott}, \citenamefont {Tram}, \citenamefont {Rampf},\ and\ \citenamefont
  {Wong}}]{Oldengott:2017fhy}%
  \BibitemOpen
  \bibfield  {author} {\bibinfo {author} {\bibfnamefont {I.~M.}\ \bibnamefont
  {Oldengott}}, \bibinfo {author} {\bibfnamefont {T.}~\bibnamefont {Tram}},
  \bibinfo {author} {\bibfnamefont {C.}~\bibnamefont {Rampf}}, \ and\ \bibinfo
  {author} {\bibfnamefont {Y.~Y.~Y.}\ \bibnamefont {Wong}},\ }\href {\doibase
  10.1088/1475-7516/2017/11/027} {\bibfield  {journal} {\bibinfo  {journal}
  {JCAP}\ }\textbf {\bibinfo {volume} {1711}},\ \bibinfo {pages} {027}
  (\bibinfo {year} {2017})},\ \Eprint {http://arxiv.org/abs/1706.02123}
  {arXiv:1706.02123 [astro-ph.CO]} \BibitemShut {NoStop}%
\bibitem [{\citenamefont {Lancaster}\ \emph {et~al.}(2017)\citenamefont
  {Lancaster}, \citenamefont {Cyr-Racine}, \citenamefont {Knox},\ and\
  \citenamefont {Pan}}]{Lancaster:2017ksf}%
  \BibitemOpen
  \bibfield  {author} {\bibinfo {author} {\bibfnamefont {L.}~\bibnamefont
  {Lancaster}}, \bibinfo {author} {\bibfnamefont {F.-Y.}\ \bibnamefont
  {Cyr-Racine}}, \bibinfo {author} {\bibfnamefont {L.}~\bibnamefont {Knox}}, \
  and\ \bibinfo {author} {\bibfnamefont {Z.}~\bibnamefont {Pan}},\ }\href
  {\doibase 10.1088/1475-7516/2017/07/033} {\bibfield  {journal} {\bibinfo
  {journal} {JCAP}\ }\textbf {\bibinfo {volume} {1707}},\ \bibinfo {pages}
  {033} (\bibinfo {year} {2017})},\ \Eprint {http://arxiv.org/abs/1704.06657}
  {arXiv:1704.06657 [astro-ph.CO]} \BibitemShut {NoStop}%
\bibitem [{\citenamefont {Kreisch}\ \emph {et~al.}(2019)\citenamefont
  {Kreisch}, \citenamefont {Cyr-Racine},\ and\ \citenamefont
  {Dor{\'e}}}]{Kreisch:2019yzn}%
  \BibitemOpen
  \bibfield  {author} {\bibinfo {author} {\bibfnamefont {C.~D.}\ \bibnamefont
  {Kreisch}}, \bibinfo {author} {\bibfnamefont {F.-Y.}\ \bibnamefont
  {Cyr-Racine}}, \ and\ \bibinfo {author} {\bibfnamefont {O.}~\bibnamefont
  {Dor{\'e}}},\ }\href@noop {} {\  (\bibinfo {year} {2019})},\ \Eprint
  {http://arxiv.org/abs/1902.00534} {arXiv:1902.00534 [astro-ph.CO]}
  \BibitemShut {NoStop}%
\bibitem [{\citenamefont {Barenboim}\ \emph {et~al.}(2019)\citenamefont
  {Barenboim}, \citenamefont {Denton},\ and\ \citenamefont
  {Oldengott}}]{Barenboim:2019tux}%
  \BibitemOpen
  \bibfield  {author} {\bibinfo {author} {\bibfnamefont {G.}~\bibnamefont
  {Barenboim}}, \bibinfo {author} {\bibfnamefont {P.~B.}\ \bibnamefont
  {Denton}}, \ and\ \bibinfo {author} {\bibfnamefont {I.~M.}\ \bibnamefont
  {Oldengott}},\ }\href {\doibase 10.1103/PhysRevD.99.083515} {\bibfield
  {journal} {\bibinfo  {journal} {Phys. Rev.}\ }\textbf {\bibinfo {volume}
  {D99}},\ \bibinfo {pages} {083515} (\bibinfo {year} {2019})},\ \Eprint
  {http://arxiv.org/abs/1903.02036} {arXiv:1903.02036 [astro-ph.CO]}
  \BibitemShut {NoStop}%
\bibitem [{\citenamefont {Archidiacono}\ and\ \citenamefont
  {Hannestad}(2014{\natexlab{b}})}]{Archidiacono:2013dua}%
  \BibitemOpen
  \bibfield  {author} {\bibinfo {author} {\bibfnamefont {M.}~\bibnamefont
  {Archidiacono}}\ and\ \bibinfo {author} {\bibfnamefont {S.}~\bibnamefont
  {Hannestad}},\ }\href {\doibase 10.1088/1475-7516/2014/07/046} {\bibfield
  {journal} {\bibinfo  {journal} {JCAP}\ }\textbf {\bibinfo {volume} {1407}},\
  \bibinfo {pages} {046} (\bibinfo {year} {2014}{\natexlab{b}})},\ \Eprint
  {http://arxiv.org/abs/1311.3873} {arXiv:1311.3873 [astro-ph.CO]} \BibitemShut
  {NoStop}%
\bibitem [{\citenamefont {Archidiacono}\ \emph {et~al.}(2015)\citenamefont
  {Archidiacono}, \citenamefont {Hannestad}, \citenamefont {Hansen},\ and\
  \citenamefont {Tram}}]{Archidiacono:2014nda}%
  \BibitemOpen
  \bibfield  {author} {\bibinfo {author} {\bibfnamefont {M.}~\bibnamefont
  {Archidiacono}}, \bibinfo {author} {\bibfnamefont {S.}~\bibnamefont
  {Hannestad}}, \bibinfo {author} {\bibfnamefont {R.~S.}\ \bibnamefont
  {Hansen}}, \ and\ \bibinfo {author} {\bibfnamefont {T.}~\bibnamefont
  {Tram}},\ }\href {\doibase 10.1103/PhysRevD.91.065021} {\bibfield  {journal}
  {\bibinfo  {journal} {Phys. Rev.}\ }\textbf {\bibinfo {volume} {D91}},\
  \bibinfo {pages} {065021} (\bibinfo {year} {2015})},\ \Eprint
  {http://arxiv.org/abs/1404.5915} {arXiv:1404.5915 [astro-ph.CO]} \BibitemShut
  {NoStop}%
\bibitem [{\citenamefont {Archidiacono}\ \emph
  {et~al.}(2016{\natexlab{a}})\citenamefont {Archidiacono}, \citenamefont
  {Gariazzo}, \citenamefont {Giunti}, \citenamefont {Hannestad}, \citenamefont
  {Hansen}, \citenamefont {Laveder},\ and\ \citenamefont
  {Tram}}]{Archidiacono:2016kkh}%
  \BibitemOpen
  \bibfield  {author} {\bibinfo {author} {\bibfnamefont {M.}~\bibnamefont
  {Archidiacono}}, \bibinfo {author} {\bibfnamefont {S.}~\bibnamefont
  {Gariazzo}}, \bibinfo {author} {\bibfnamefont {C.}~\bibnamefont {Giunti}},
  \bibinfo {author} {\bibfnamefont {S.}~\bibnamefont {Hannestad}}, \bibinfo
  {author} {\bibfnamefont {R.}~\bibnamefont {Hansen}}, \bibinfo {author}
  {\bibfnamefont {M.}~\bibnamefont {Laveder}}, \ and\ \bibinfo {author}
  {\bibfnamefont {T.}~\bibnamefont {Tram}},\ }\href {\doibase
  10.1088/1475-7516/2016/08/067} {\bibfield  {journal} {\bibinfo  {journal}
  {JCAP}\ }\textbf {\bibinfo {volume} {1608}},\ \bibinfo {pages} {067}
  (\bibinfo {year} {2016}{\natexlab{a}})},\ \Eprint
  {http://arxiv.org/abs/1606.07673} {arXiv:1606.07673 [astro-ph.CO]}
  \BibitemShut {NoStop}%
\bibitem [{\citenamefont {Archidiacono}\ \emph
  {et~al.}(2016{\natexlab{b}})\citenamefont {Archidiacono}, \citenamefont
  {Hannestad}, \citenamefont {Hansen},\ and\ \citenamefont
  {Tram}}]{Archidiacono:2015oma}%
  \BibitemOpen
  \bibfield  {author} {\bibinfo {author} {\bibfnamefont {M.}~\bibnamefont
  {Archidiacono}}, \bibinfo {author} {\bibfnamefont {S.}~\bibnamefont
  {Hannestad}}, \bibinfo {author} {\bibfnamefont {R.~S.}\ \bibnamefont
  {Hansen}}, \ and\ \bibinfo {author} {\bibfnamefont {T.}~\bibnamefont
  {Tram}},\ }\href {\doibase 10.1103/PhysRevD.93.045004} {\bibfield  {journal}
  {\bibinfo  {journal} {Phys. Rev.}\ }\textbf {\bibinfo {volume} {D93}},\
  \bibinfo {pages} {045004} (\bibinfo {year} {2016}{\natexlab{b}})},\ \Eprint
  {http://arxiv.org/abs/1508.02504} {arXiv:1508.02504 [astro-ph.CO]}
  \BibitemShut {NoStop}%
\bibitem [{\citenamefont {Chu}\ \emph {et~al.}(2015)\citenamefont {Chu},
  \citenamefont {Dasgupta},\ and\ \citenamefont {Kopp}}]{Chu:2015ipa}%
  \BibitemOpen
  \bibfield  {author} {\bibinfo {author} {\bibfnamefont {X.}~\bibnamefont
  {Chu}}, \bibinfo {author} {\bibfnamefont {B.}~\bibnamefont {Dasgupta}}, \
  and\ \bibinfo {author} {\bibfnamefont {J.}~\bibnamefont {Kopp}},\ }\href
  {\doibase 10.1088/1475-7516/2015/10/011} {\bibfield  {journal} {\bibinfo
  {journal} {JCAP}\ }\textbf {\bibinfo {volume} {1510}},\ \bibinfo {pages}
  {011} (\bibinfo {year} {2015})},\ \Eprint {http://arxiv.org/abs/1505.02795}
  {arXiv:1505.02795 [hep-ph]} \BibitemShut {NoStop}%
\bibitem [{\citenamefont {Song}\ \emph {et~al.}(2018)\citenamefont {Song},
  \citenamefont {Gonzalez-Garcia},\ and\ \citenamefont
  {Salvado}}]{Song:2018zyl}%
  \BibitemOpen
  \bibfield  {author} {\bibinfo {author} {\bibfnamefont {N.}~\bibnamefont
  {Song}}, \bibinfo {author} {\bibfnamefont {M.~C.}\ \bibnamefont
  {Gonzalez-Garcia}}, \ and\ \bibinfo {author} {\bibfnamefont {J.}~\bibnamefont
  {Salvado}},\ }\href {\doibase 10.1088/1475-7516/2018/10/055} {\bibfield
  {journal} {\bibinfo  {journal} {JCAP}\ }\textbf {\bibinfo {volume} {1810}},\
  \bibinfo {pages} {055} (\bibinfo {year} {2018})},\ \Eprint
  {http://arxiv.org/abs/1805.08218} {arXiv:1805.08218 [astro-ph.CO]}
  \BibitemShut {NoStop}%
\bibitem [{\citenamefont {Oldengott}\ \emph
  {et~al.}(2015{\natexlab{b}})\citenamefont {Oldengott}, \citenamefont
  {Rampf},\ and\ \citenamefont {Wong}}]{Oldengott:2014qra}%
  \BibitemOpen
  \bibfield  {author} {\bibinfo {author} {\bibfnamefont {I.~M.}\ \bibnamefont
  {Oldengott}}, \bibinfo {author} {\bibfnamefont {C.}~\bibnamefont {Rampf}}, \
  and\ \bibinfo {author} {\bibfnamefont {Y.~Y.~Y.}\ \bibnamefont {Wong}},\
  }\href {\doibase 10.1088/1475-7516/2015/04/016} {\bibfield  {journal}
  {\bibinfo  {journal} {JCAP}\ }\textbf {\bibinfo {volume} {1504}},\ \bibinfo
  {pages} {016} (\bibinfo {year} {2015}{\natexlab{b}})},\ \Eprint
  {http://arxiv.org/abs/1409.1577} {arXiv:1409.1577 [astro-ph.CO]} \BibitemShut
  {NoStop}%
\bibitem [{\citenamefont {Ma}\ and\ \citenamefont
  {Bertschinger}(1994)}]{Ma:1994dv}%
  \BibitemOpen
  \bibfield  {author} {\bibinfo {author} {\bibfnamefont {C.-P.}\ \bibnamefont
  {Ma}}\ and\ \bibinfo {author} {\bibfnamefont {E.}~\bibnamefont
  {Bertschinger}},\ }\href@noop {} {\bibfield  {journal} {\bibinfo  {journal}
  {Submitted to: Astrophys. J.}\ } (\bibinfo {year} {1994})},\ \Eprint
  {http://arxiv.org/abs/astro-ph/9401007} {arXiv:astro-ph/9401007 [astro-ph]}
  \BibitemShut {NoStop}%
\bibitem [{\citenamefont {Hannestad}(2005)}]{Hannestad:2004qu}%
  \BibitemOpen
  \bibfield  {author} {\bibinfo {author} {\bibfnamefont {S.}~\bibnamefont
  {Hannestad}},\ }\href {\doibase 10.1088/1475-7516/2005/02/011} {\bibfield
  {journal} {\bibinfo  {journal} {JCAP}\ }\textbf {\bibinfo {volume} {0502}},\
  \bibinfo {pages} {011} (\bibinfo {year} {2005})},\ \Eprint
  {http://arxiv.org/abs/astro-ph/0411475} {arXiv:astro-ph/0411475 [astro-ph]}
  \BibitemShut {NoStop}%
\bibitem [{\citenamefont {Adam}\ \emph {et~al.}(2016)\citenamefont {Adam} \emph
  {et~al.}}]{Adam:2015rua}%
  \BibitemOpen
  \bibfield  {author} {\bibinfo {author} {\bibfnamefont {R.}~\bibnamefont
  {Adam}} \emph {et~al.} (\bibinfo {collaboration} {Planck}),\ }\href {\doibase
  10.1051/0004-6361/201527101} {\bibfield  {journal} {\bibinfo  {journal}
  {Astron. Astrophys.}\ }\textbf {\bibinfo {volume} {594}},\ \bibinfo {pages}
  {A1} (\bibinfo {year} {2016})},\ \Eprint {http://arxiv.org/abs/1502.01582}
  {arXiv:1502.01582 [astro-ph.CO]} \BibitemShut {NoStop}%
\bibitem [{\citenamefont {Aghanim}\ \emph {et~al.}(2016)\citenamefont {Aghanim}
  \emph {et~al.}}]{Aghanim:2015xee}%
  \BibitemOpen
  \bibfield  {author} {\bibinfo {author} {\bibfnamefont {N.}~\bibnamefont
  {Aghanim}} \emph {et~al.} (\bibinfo {collaboration} {Planck}),\ }\href
  {\doibase 10.1051/0004-6361/201526926} {\bibfield  {journal} {\bibinfo
  {journal} {Astron. Astrophys.}\ }\textbf {\bibinfo {volume} {594}},\ \bibinfo
  {pages} {A11} (\bibinfo {year} {2016})},\ \Eprint
  {http://arxiv.org/abs/1507.02704} {arXiv:1507.02704 [astro-ph.CO]}
  \BibitemShut {NoStop}%
\bibitem [{\citenamefont {Ade}\ \emph {et~al.}(2016{\natexlab{a}})\citenamefont
  {Ade} \emph {et~al.}}]{Ade:2015zua}%
  \BibitemOpen
  \bibfield  {author} {\bibinfo {author} {\bibfnamefont {P.~A.~R.}\
  \bibnamefont {Ade}} \emph {et~al.} (\bibinfo {collaboration} {Planck}),\
  }\href {\doibase 10.1051/0004-6361/201525941} {\bibfield  {journal} {\bibinfo
   {journal} {Astron. Astrophys.}\ }\textbf {\bibinfo {volume} {594}},\
  \bibinfo {pages} {A15} (\bibinfo {year} {2016}{\natexlab{a}})},\ \Eprint
  {http://arxiv.org/abs/1502.01591} {arXiv:1502.01591 [astro-ph.CO]}
  \BibitemShut {NoStop}%
\bibitem [{\citenamefont {Beutler}\ \emph {et~al.}(2011)\citenamefont
  {Beutler}, \citenamefont {Blake}, \citenamefont {Colless}, \citenamefont
  {Jones}, \citenamefont {Staveley-Smith}, \citenamefont {Campbell},
  \citenamefont {Parker}, \citenamefont {Saunders},\ and\ \citenamefont
  {Watson}}]{Beutler:2011hx}%
  \BibitemOpen
  \bibfield  {author} {\bibinfo {author} {\bibfnamefont {F.}~\bibnamefont
  {Beutler}}, \bibinfo {author} {\bibfnamefont {C.}~\bibnamefont {Blake}},
  \bibinfo {author} {\bibfnamefont {M.}~\bibnamefont {Colless}}, \bibinfo
  {author} {\bibfnamefont {D.~H.}\ \bibnamefont {Jones}}, \bibinfo {author}
  {\bibfnamefont {L.}~\bibnamefont {Staveley-Smith}}, \bibinfo {author}
  {\bibfnamefont {L.}~\bibnamefont {Campbell}}, \bibinfo {author}
  {\bibfnamefont {Q.}~\bibnamefont {Parker}}, \bibinfo {author} {\bibfnamefont
  {W.}~\bibnamefont {Saunders}}, \ and\ \bibinfo {author} {\bibfnamefont
  {F.}~\bibnamefont {Watson}},\ }\href {\doibase
  10.1111/j.1365-2966.2011.19250.x} {\bibfield  {journal} {\bibinfo  {journal}
  {Mon. Not. Roy. Astron. Soc.}\ }\textbf {\bibinfo {volume} {416}},\ \bibinfo
  {pages} {3017} (\bibinfo {year} {2011})},\ \Eprint
  {http://arxiv.org/abs/1106.3366} {arXiv:1106.3366 [astro-ph.CO]} \BibitemShut
  {NoStop}%
\bibitem [{\citenamefont {Alam}\ \emph {et~al.}(2017)\citenamefont {Alam} \emph
  {et~al.}}]{Alam:2016hwk}%
  \BibitemOpen
  \bibfield  {author} {\bibinfo {author} {\bibfnamefont {S.}~\bibnamefont
  {Alam}} \emph {et~al.} (\bibinfo {collaboration} {BOSS}),\ }\href {\doibase
  10.1093/mnras/stx721} {\bibfield  {journal} {\bibinfo  {journal} {Mon. Not.
  Roy. Astron. Soc.}\ }\textbf {\bibinfo {volume} {470}},\ \bibinfo {pages}
  {2617} (\bibinfo {year} {2017})},\ \Eprint {http://arxiv.org/abs/1607.03155}
  {arXiv:1607.03155 [astro-ph.CO]} \BibitemShut {NoStop}%
\bibitem [{\citenamefont {Anderson}\ \emph {et~al.}(2014)\citenamefont
  {Anderson}, \citenamefont {Aubourg}, \citenamefont {Bailey}, \citenamefont
  {Beutler}, \citenamefont {Bhardwaj}, \citenamefont {Blanton}, \citenamefont
  {Bolton}, \citenamefont {Brinkmann}, \citenamefont {Brownstein},
  \citenamefont {Burden} \emph {et~al.}}]{anderson2014clustering}%
  \BibitemOpen
  \bibfield  {author} {\bibinfo {author} {\bibfnamefont {L.}~\bibnamefont
  {Anderson}}, \bibinfo {author} {\bibfnamefont {{\'E}.}~\bibnamefont
  {Aubourg}}, \bibinfo {author} {\bibfnamefont {S.}~\bibnamefont {Bailey}},
  \bibinfo {author} {\bibfnamefont {F.}~\bibnamefont {Beutler}}, \bibinfo
  {author} {\bibfnamefont {V.}~\bibnamefont {Bhardwaj}}, \bibinfo {author}
  {\bibfnamefont {M.}~\bibnamefont {Blanton}}, \bibinfo {author} {\bibfnamefont
  {A.~S.}\ \bibnamefont {Bolton}}, \bibinfo {author} {\bibfnamefont
  {J.}~\bibnamefont {Brinkmann}}, \bibinfo {author} {\bibfnamefont {J.~R.}\
  \bibnamefont {Brownstein}}, \bibinfo {author} {\bibfnamefont
  {A.}~\bibnamefont {Burden}},  \emph {et~al.},\ }\href@noop {} {\bibfield
  {journal} {\bibinfo  {journal} {Monthly Notices of the Royal Astronomical
  Society}\ }\textbf {\bibinfo {volume} {441}},\ \bibinfo {pages} {24}
  (\bibinfo {year} {2014})}\BibitemShut {NoStop}%
\bibitem [{\citenamefont {Betoule}\ \emph {et~al.}(2013)\citenamefont {Betoule}
  \emph {et~al.}}]{Betoule:2012an}%
  \BibitemOpen
  \bibfield  {author} {\bibinfo {author} {\bibfnamefont {M.}~\bibnamefont
  {Betoule}} \emph {et~al.} (\bibinfo {collaboration} {SDSS}),\ }\href
  {\doibase 10.1051/0004-6361/201220610} {\bibfield  {journal} {\bibinfo
  {journal} {Astron. Astrophys.}\ }\textbf {\bibinfo {volume} {552}},\ \bibinfo
  {pages} {A124} (\bibinfo {year} {2013})},\ \Eprint
  {http://arxiv.org/abs/1212.4864} {arXiv:1212.4864 [astro-ph.CO]} \BibitemShut
  {NoStop}%
\bibitem [{\citenamefont {Efstathiou}(2014)}]{Efstathiou:2013via}%
  \BibitemOpen
  \bibfield  {author} {\bibinfo {author} {\bibfnamefont {G.}~\bibnamefont
  {Efstathiou}},\ }\href {\doibase 10.1093/mnras/stu278} {\bibfield  {journal}
  {\bibinfo  {journal} {Mon. Not. Roy. Astron. Soc.}\ }\textbf {\bibinfo
  {volume} {440}},\ \bibinfo {pages} {1138} (\bibinfo {year} {2014})},\ \Eprint
  {http://arxiv.org/abs/1311.3461} {arXiv:1311.3461 [astro-ph.CO]} \BibitemShut
  {NoStop}%
\bibitem [{\citenamefont {Lewis}\ \emph {et~al.}(2000)\citenamefont {Lewis},
  \citenamefont {Challinor},\ and\ \citenamefont {Lasenby}}]{Lewis:1999bs}%
  \BibitemOpen
  \bibfield  {author} {\bibinfo {author} {\bibfnamefont {A.}~\bibnamefont
  {Lewis}}, \bibinfo {author} {\bibfnamefont {A.}~\bibnamefont {Challinor}}, \
  and\ \bibinfo {author} {\bibfnamefont {A.}~\bibnamefont {Lasenby}},\ }\href
  {\doibase 10.1086/309179} {\bibfield  {journal} {\bibinfo  {journal}
  {Astrophys. J.}\ }\textbf {\bibinfo {volume} {538}},\ \bibinfo {pages} {473}
  (\bibinfo {year} {2000})},\ \Eprint {http://arxiv.org/abs/astro-ph/9911177}
  {arXiv:astro-ph/9911177 [astro-ph]} \BibitemShut {NoStop}%
\bibitem [{\citenamefont {Ade}\ \emph {et~al.}(2016{\natexlab{b}})\citenamefont
  {Ade} \emph {et~al.}}]{Ade:2015xua}%
  \BibitemOpen
  \bibfield  {author} {\bibinfo {author} {\bibfnamefont {P.~A.~R.}\
  \bibnamefont {Ade}} \emph {et~al.} (\bibinfo {collaboration} {Planck}),\
  }\href {\doibase 10.1051/0004-6361/201525830} {\bibfield  {journal} {\bibinfo
   {journal} {Astron. Astrophys.}\ }\textbf {\bibinfo {volume} {594}},\
  \bibinfo {pages} {A13} (\bibinfo {year} {2016}{\natexlab{b}})},\ \Eprint
  {http://arxiv.org/abs/1502.01589} {arXiv:1502.01589 [astro-ph.CO]}
  \BibitemShut {NoStop}%
\bibitem [{\citenamefont {Percival}\ \emph {et~al.}(2002)\citenamefont
  {Percival} \emph {et~al.}}]{Percival:2002gq}%
  \BibitemOpen
  \bibfield  {author} {\bibinfo {author} {\bibfnamefont {W.~J.}\ \bibnamefont
  {Percival}} \emph {et~al.} (\bibinfo {collaboration} {2dFGRS Team}),\ }\href
  {\doibase 10.1046/j.1365-8711.2002.06001.x} {\bibfield  {journal} {\bibinfo
  {journal} {Mon. Not. Roy. Astron. Soc.}\ }\textbf {\bibinfo {volume} {337}},\
  \bibinfo {pages} {1068} (\bibinfo {year} {2002})},\ \Eprint
  {http://arxiv.org/abs/astro-ph/0206256} {arXiv:astro-ph/0206256 [astro-ph]}
  \BibitemShut {NoStop}%
\bibitem [{\citenamefont {Galli}\ \emph {et~al.}(2014)\citenamefont {Galli},
  \citenamefont {Benabed}, \citenamefont {Bouchet}, \citenamefont {Cardoso},
  \citenamefont {Elsner}, \citenamefont {Hivon}, \citenamefont {Mangilli},
  \citenamefont {Prunet},\ and\ \citenamefont {Wandelt}}]{Galli:2014kla}%
  \BibitemOpen
  \bibfield  {author} {\bibinfo {author} {\bibfnamefont {S.}~\bibnamefont
  {Galli}}, \bibinfo {author} {\bibfnamefont {K.}~\bibnamefont {Benabed}},
  \bibinfo {author} {\bibfnamefont {F.}~\bibnamefont {Bouchet}}, \bibinfo
  {author} {\bibfnamefont {J.-F.}\ \bibnamefont {Cardoso}}, \bibinfo {author}
  {\bibfnamefont {F.}~\bibnamefont {Elsner}}, \bibinfo {author} {\bibfnamefont
  {E.}~\bibnamefont {Hivon}}, \bibinfo {author} {\bibfnamefont
  {A.}~\bibnamefont {Mangilli}}, \bibinfo {author} {\bibfnamefont
  {S.}~\bibnamefont {Prunet}}, \ and\ \bibinfo {author} {\bibfnamefont
  {B.}~\bibnamefont {Wandelt}},\ }\href {\doibase 10.1103/PhysRevD.90.063504}
  {\bibfield  {journal} {\bibinfo  {journal} {Phys. Rev.}\ }\textbf {\bibinfo
  {volume} {D90}},\ \bibinfo {pages} {063504} (\bibinfo {year} {2014})},\
  \Eprint {http://arxiv.org/abs/1403.5271} {arXiv:1403.5271 [astro-ph.CO]}
  \BibitemShut {NoStop}%
\bibitem [{\citenamefont {Riess}\ \emph {et~al.}(2016)\citenamefont {Riess}
  \emph {et~al.}}]{Riess:2016jrr}%
  \BibitemOpen
  \bibfield  {author} {\bibinfo {author} {\bibfnamefont {A.~G.}\ \bibnamefont
  {Riess}} \emph {et~al.},\ }\href {\doibase 10.3847/0004-637X/826/1/56}
  {\bibfield  {journal} {\bibinfo  {journal} {Astrophys. J.}\ }\textbf
  {\bibinfo {volume} {826}},\ \bibinfo {pages} {56} (\bibinfo {year} {2016})},\
  \Eprint {http://arxiv.org/abs/1604.01424} {arXiv:1604.01424 [astro-ph.CO]}
  \BibitemShut {NoStop}%
\bibitem [{\citenamefont {Srednicki}(2007)}]{Srednicki:2007qs}%
  \BibitemOpen
  \bibfield  {author} {\bibinfo {author} {\bibfnamefont {M.}~\bibnamefont
  {Srednicki}},\ }\href@noop {} {\emph {\bibinfo {title} {{Quantum field
  theory}}}}\ (\bibinfo  {publisher} {Cambridge University Press},\ \bibinfo
  {year} {2007})\BibitemShut {NoStop}%
\bibitem [{\citenamefont {Dreiner}\ \emph {et~al.}(2010)\citenamefont
  {Dreiner}, \citenamefont {Haber},\ and\ \citenamefont
  {Martin}}]{Dreiner:2008tw}%
  \BibitemOpen
  \bibfield  {author} {\bibinfo {author} {\bibfnamefont {H.~K.}\ \bibnamefont
  {Dreiner}}, \bibinfo {author} {\bibfnamefont {H.~E.}\ \bibnamefont {Haber}},
  \ and\ \bibinfo {author} {\bibfnamefont {S.~P.}\ \bibnamefont {Martin}},\
  }\href {\doibase 10.1016/j.physrep.2010.05.002} {\bibfield  {journal}
  {\bibinfo  {journal} {Phys. Rept.}\ }\textbf {\bibinfo {volume} {494}},\
  \bibinfo {pages} {1} (\bibinfo {year} {2010})},\ \Eprint
  {http://arxiv.org/abs/0812.1594} {arXiv:0812.1594 [hep-ph]} \BibitemShut
  {NoStop}%
\bibitem [{\citenamefont {Kolb}\ and\ \citenamefont
  {Turner}(1990)}]{Kolb:1990vq}%
  \BibitemOpen
  \bibfield  {author} {\bibinfo {author} {\bibfnamefont {E.~W.}\ \bibnamefont
  {Kolb}}\ and\ \bibinfo {author} {\bibfnamefont {M.~S.}\ \bibnamefont
  {Turner}},\ }\href@noop {} {\bibfield  {journal} {\bibinfo  {journal} {Front.
  Phys.}\ }\textbf {\bibinfo {volume} {69}},\ \bibinfo {pages} {1} (\bibinfo
  {year} {1990})}\BibitemShut {NoStop}%
\end{thebibliography}%

\end{document}